\documentclass[journal=jctcce,manuscript=article]{achemso}

\usepackage{amsmath, amssymb, amsfonts}
\usepackage{array, booktabs, tabularx}
\usepackage{xcolor, graphicx, tikz}
\usepackage[caption=false]{subfig}
\usepackage{hyperref}  
\usepackage{braket}
\usepackage{bm}
\usepackage{cancel}
\usepackage[ruled]{algorithm2e}
\usepackage[normalem]{ulem} 

\hypersetup{colorlinks = True, allcolors = {blue}}

\SetAlCapSty{textnormal} 

\newcommand{\KetBra}[2]{ \ket{#1} \bra{#2} }

\newcommand{\FC}[1]{ f_{#1}^\dagger }
\newcommand{\FA}[1]{ f_{#1} }

\newcommand{\Pauli}[2]{ \mathbb{P}_{#2}^{(#1)} }

\newcommand{\SP}[1]{ \sigma^{+}_{#1} }
\newcommand{\SM}[1]{ \sigma^{-}_{#1} }

\newcommand{\BC}[1]{ b_{#1}^\dagger }
\newcommand{\BA}[1]{ b_{#1} }
\newcommand{\BX}[1]{ \hat{x}_{#1} }
\newcommand{\BP}[1]{ \hat{p}_{#1} }

\newcommand{\UniAn}[1]{ U_A^{(#1)} }
\newcommand{\UniGate}[2]{ U_{#2}^{(#1)} }

\def\UniSNAP{ U_{SD} }

\def\HElec{ H_{\text{elec}} }

\def\HamQ{ H_Q }
\newcommand{\HOne}[2]{ h_{#2}^{#1} }
\newcommand{\HTwo}[2]{ v_{#2}^{#1} }
\newcommand{\HQCoeff}[1]{ g_{#1} }

\def\EYE{ \mathbb{I} }

\newcommand{\ComCom}[1]{\mathcal{O}(#1)}

\newcommand{\Eq}[1]{Eq.~({#1})}
\newcommand{\Fig}[1]{Figure~{#1}}

\newcommand{\Sec}[1]{Section~{#1}}

\newcommand{\Reference}[1]{Ref.~{#1}}

\title{Qumode-Based Variational Quantum Eigensolver for Molecular Excited States}

\author{Rishab Dutta}
\affiliation{Department of Chemistry, Yale University, New Haven, Connecticut 06520, USA}

\author{Cameron Cianci}
\affiliation{Department of Physics, University of Connecticut, Storrs, Connecticut 06269, USA}
\affiliation{Mirion Technologies (Canberra) Inc., 800 Research Parkway, Meriden, Connecticut 06450, USA}

\author{Alexander V. Soudackov}
\affiliation{Department of Chemistry, Yale University, New Haven, Connecticut 06520, USA}

\author{Yuchen Wang}
\affiliation{Department of Chemistry and The James Franck Institute, The University of Chicago, Chicago, Illinois 60637, USA}

\author{Chuzhi Xu}
\affiliation{Department of Chemistry, Yale University, New Haven, Connecticut 06520, USA}

\author{David A. Mazziotti}
\affiliation{Department of Chemistry and The James Franck Institute, The University of Chicago, Chicago, Illinois 60637, USA}

\author{Lea F. Santos}
\affiliation{Department of Physics, University of Connecticut, Storrs, Connecticut 06269, USA}

\author{Victor S. Batista}
\email{victor.batista@yale.edu}
\affiliation{Department of Chemistry, Yale University, New Haven, Connecticut 06520, USA}
\affiliation{Yale Quantum Institute, Yale University, New Haven, Connecticut 06511, USA}

\begin{document}


\begin{abstract}

We introduce the \textit{Qumode Subspace Variational Quantum Eigensolver} (QSS-VQE), a hybrid quantum-classical algorithm for computing molecular excited states using the Fock basis of bosonic qumodes in circuit quantum electrodynamics (cQED) devices. 
This approach harnesses the native universal gate sets of qubit-qumode architectures to construct highly expressive variational ansatze, offering potential advantages over conventional qubit-based methods. In QSS-VQE, the electronic structure Hamiltonian is first mapped to a qubit representation and subsequently embedded into the Fock space of bosonic qumodes, enabling efficient state preparation and reduced quantum resource requirements. We demonstrate the performance of QSS-VQE through simulations of molecular excited states, including dihydrogen and a conical intersection in cytosine. Additionally, we explore a bosonic model Hamiltonian to assess the expressivity of qumode gates, identifying regimes where qumode-based implementations outperform purely qubit-based approaches. These results highlight the promise of leveraging bosonic degrees of freedom for enhanced quantum simulation of complex molecular systems.

\end{abstract}


\maketitle


\section{Introduction} \label{sec: intro}

The electronic structure problem is a central challenge in quantum chemistry and materials science, driven by the exponential growth of the Hilbert space with the number of molecular orbitals.~\cite{SzaboBook,HelgakerBook} 
This scaling quickly makes exact solutions infeasible for classical algorithms as system size increases. Quantum computers offer a promising alternative, as they naturally operate within exponentially large Hilbert spaces and can, in principle, represent many-body quantum states more efficiently.~\cite{Feynman1982,McArdle2020} To take advantage of early quantum hardware, particularly noisy intermediate-scale quantum (NISQ) devices, hybrid quantum-classical approaches have been proposed. Among these, the variational quantum eigensolver (VQE)~\cite{Peruzzo2014} has been widely used to estimate the ground state energy of electronic Hamiltonians by classically optimizing the variational parameters of an ansatz quantum circuit.

Extending ground-state VQE frameworks to compute excited states has become an active area of research,~\cite{Nakanishi_2019,McClean2017,Higgott_2019,wang2023electronicexcitedstatesvariancebased} driven by the broad importance of excited-state properties in chemical and materials applications. 
These properties are critical for tasks such as designing molecular catalysts, \cite{Reiher_2017} characterizing conical intersections and fluorescence mechanisms,\cite{Mei2019excited}
and analyzing photodissociation pathways and reactive intermediates. \cite{Higgott_2019,Lischka2018} 
Several algorithms have been developed that generalize the variational approach to multiple eigenstates. Notable examples include the subspace-search variational quantum eigensolver (SSVQE),~\cite{Nakanishi_2019} the variational quantum deflation (VQD) algorithm,~\cite{Higgott_2019} 
and the Subspace-Search Quantum Imaginary Time Evolution (SSQITE) algorithm,~\cite{cianci2024subspacesearchquantumimaginarytime} all of which optimize over multiple orthogonal states within a single variational framework. These methods, together with VQE, belong to the broader class of variational quantum algorithms (VQAs).~\cite{Mcclean2016theory}
Central to the effectiveness of any VQA is the choice of ansatz, which must simultaneously offer sufficient expressiveness and remain compatible with the constraints of near-term quantum hardware.
Recent NISQ-oriented methods for ground and excited states have also aimed to bypass explicit variational optimization. Examples of these approaches include generating a compact, non-orthogonal manifold of states and extract multiple energies through a single generalized eigenvalue problem \cite{Baek2023,Zheng2024GCM}, while others rely on moments-based expansions that similarly avoid iterative parameter tuning.\cite{Claudino2023modeling}

The advent of bosonic quantum devices has introduced a new paradigm for variational quantum algorithms.
\cite{Dutta2024EST, Dutta2025solving, Zhang2024energy,Kan2024,Liao2024quantum,Araz2025hybrid} 
Instead of encoding information in discrete qubits, these platforms leverage quantum harmonic oscillators (QHOs), or qumodes, to represent quantum states.\cite{Dutta2024perspective} Qumodes offer access to large Hilbert spaces via Fock state representations and support native gate operations that would require deep circuits to emulate on qubit-based architectures. Circuit quantum electrodynamics (cQED) provides a leading hardware realization of this model, employing microwave cavities as bosonic modes and transmon qubits to implement universal control.~\cite{Blais2021,Dutta2024perspective,Liu2024qumodequbitreview,Copetudo2024}

In this paper, we implement a variational algorithm for computing excited states of electronic Hamiltonians on hybrid qubit-qumode architectures. 
Building on our previous protocol for ground state calculations, which encodes multiple qubits into the Fock space of a single qumode,~\cite{Dutta2024EST} we generalize the approach to access excited states. Expectation values of qubit-encoded Hamiltonians are obtained by measuring the qumode photon number distribution and mapping it to computational basis outcomes. The resulting qumode subspace variational quantum eigensolver (QSS-VQE) is structured within the SSVQE framework,\cite{Nakanishi_2019} using a shared parameterized circuit applied to an orthonormal set of input states. On the qumode platform, these inputs are naturally realized as Fock states---discrete photon number levels that can be efficiently initialized using established cQED techniques.\cite{Kudra2022,Job2023efficient,Krastanov2015,Huang2025fast} This architecture enables the design of bosonic ansatze that are native to qubit–qumode devices and particularly well-suited for representing electronic excited states. Numerical benchmarks on representative molecular systems show that qumode-based circuits achieve accuracy on par with or better than conventional qubit-based ansatze. In particular, we demonstrate a case where a simple qumode gate yields a more compact and accurate representation of both ground and excited states than standard qubit circuits.

The remainder of this paper is organized as follows. Section~\ref{sec: background} reviews the electronic structure Hamiltonian, variational algorithms for excited states, and the foundations of bosonic quantum computing. Section~\ref{sec: subspace_vqe_qumode} introduces the qumode-based subspace VQE (QSS-VQE) algorithm. Section~\ref{sec: applications} presents benchmark results and highlights cases where qumode gates provide clear advantages over conventional approaches. Finally, Section~\ref{sec: final} discusses the broader implications of this work and outlines future directions.


\section{Background} \label{sec: background}

In this section, we review the electronic structure Hamiltonian and excited state methods based on qubit-based algorithms before discussing elementary concepts of bosonic quantum computing. 

\subsection{Electronic structure Hamiltonian}

The electronic Hamiltonian is represented in the molecular spin-orbital basis, as follows:~\cite{HelgakerBook} 
\begin{equation} \label{eq: molecular_ham}
\HElec
= \sum_{pq} \HOne{p}{q} \: \FC{p} \FA{q}
+ \frac{1}{2} \: \sum_{pqrs} 
\HTwo{pq}{rs} \: \FC{p} \FC{q} \FA{s} \FA{r},
\end{equation}
where the $p, q, r, s$ indices label the spin-orbitals and 
$ \{ \FC{p}, \FA{q} \} $ are fermionic creation and annihilation operators.
The scalars $ \{ \HOne{p}{q} \} $ and $ \{ \HTwo{pq}{rs} \} $ are the one-electron and two-electron integrals and can be precomputed with a Hartree--Fock calculation on a classical computer.~\cite{SzaboBook,HelgakerBook}

A typical quantum algorithm for the molecular electronic structure first converts \Eq{\ref{eq: molecular_ham}} to a qubit Hamiltonian. 
The elementary one-qubit operators are represented by Pauli matrices defined below
\begin{equation}
X 
= \begin{pmatrix}
0 & 1 \\ 1 & 0
\end{pmatrix}, \quad
Y 
= \begin{pmatrix}
0 & -i \\ i & 0
\end{pmatrix} , \quad
Z 
= \begin{pmatrix}
1 & 0 \\ 0 & -1
\end{pmatrix}.
\end{equation}
We will use the symbols $\{ \sigma_j \}$ with $j \in \{ x, y, z\}$ for operators and 
$\{ X, Y, Z \}$ for Pauli matrices interchangeably.
Pauli words are defined as the tensor products of Pauli matrices.
For example, 
$ X_1 \: Y_2 = X \otimes Y $ represents a two-qubit operator, where $\otimes$ denotes tensor product. 
The molecular electronic Hamiltonian in \Eq{\ref{eq: molecular_ham}} can be transformed to a qubit Hamiltonian of the form in \Eq{\ref{eq: qubit_ham}} by applying a fermion to qubit mapping
\begin{equation} \label{eq: qubit_ham}
\HamQ 
= \sum_{j = 1}^{N_H} \HQCoeff{j} \: 
\bigotimes_{\mu = 1}^{N_Q} \: \sigma_{j, \mu} 
= \sum_{j = 1}^{N_H} \: \HQCoeff{j} \: \Pauli{N_Q}{j}, 
\end{equation}
where $\{ \HQCoeff{\mu} \}$ are scalar coefficients and $N_Q$ is the number of qubits. 
One of the most well-known fermion-to-qubit mappings is the Jordan--Wigner transformation (JWT),~\cite{Jordan1935} 
\begin{subequations}
\begin{align}
\FC{p}
&\mapsto \frac{1}{2} \: ( X_p - i Y_p ) \: 
\bigotimes_{q < p} \: Z_q,
\\
\FA{p}
&\mapsto \frac{1}{2} \: ( X_p + i Y_p ) \: 
\bigotimes_{q < p} \: Z_q,
\end{align}    
\end{subequations}
where the qubit indices represent the spin-orbital indices of the fermionic operators. The
JWT maps the electronic Hamiltonian to \Eq{\ref{eq: qubit_ham}} with $N_H = \ComCom{M^4}$ and $N_Q = M$, where $M$ is the number of spin-orbitals. 

\subsection{Excited state quantum algorithms on conventional qubit-based hardware} 

The variational quantum eigensolver (VQE) is a hybrid quantum-classical algorithm commonly used to obtain the ground electronic state energy of a molecule. A quantum device generates a trial statevector and the expectation values of Pauli words necessary to compute the energy expectation value:
\begin{equation} \label{eq: vqe}
\min_{\bm{\theta}} E 
= \frac{ \braket{ \psi (\bm{\theta}) | \HamQ | \psi (\bm{\theta}) } }{ \braket{ \psi (\bm{\theta}) | \psi (\bm{\theta}) } }
= \frac{ \braket{ \bm{0} | U^\dagger (\bm{\theta}) \: 
\HamQ \: U (\bm{\theta}) | \bm{0} } }{ \braket{ \bm{0} | U^\dagger (\bm{\theta}) \: U (\bm{\theta}) | \bm{0} } }
= \braket{ \bm{0} | U^\dagger (\bm{\theta}) \: 
\HamQ \: U (\bm{\theta}) | \bm{0} },
\end{equation}
while the device variational parameters $\{ \theta_j \}$ are updated by an optimizer on a classical computer.  
The state $\ket{\bm{0}}$ in \Eq{\ref{eq: vqe}} represents an easily prepared initial multi-qubit state, such as the vacuum state with all qubits in $\ket{0}$.
$ U (\bm{\theta}) $ represents a multi-qubit parameterized quantum circuit defining the set of unitary pulses applied to the quantum register.

Several algorithms have been developed to extend the Variational Quantum Eigensolver (VQE) to excited-state calculations. The earliest of these is the Variational Quantum Deflation (VQD) method~\cite{Higgott_2019}, which augments the VQE cost function with orthogonality penalties to previously obtained eigenstates. This enables sequential targeting of low-lying excited states.
Other ground-state eigensolvers have also been adapted for excited-state estimation. Notably, quantum imaginary time evolution (QITE)~\cite{McArdle_2019} has been extended to access excited-state manifolds through subspace methods and projection techniques.~\cite{cianci2024subspacesearchquantumimaginarytime,doi:10.1021/acs.jctc.2c00906} Similarly, the contracted quantum eigensolver (CQE)~\cite{Smart2021,wang2023electronicexcitedstatesvariancebased,Smart_2024,Benavides_Riveros_2024} has been generalized to support excited-state calculations by incorporating state-selective constraints and variance reduction strategies.
In addition, quantum subspace expansion methods have been applied to calculations of excited states on quantum devices by recasting the problem into a small basis of non-orthogonal states and solving the generalized eigenvalue problem by optimization on this smaller subspace.~\cite{Colless2018spectra, Motta_2019} These methods, including QLanczos~\cite{Motta_2019} and QDavidson,~\cite{tkachenko2023quantumdavidsonalgorithmexcited} that often use quantum imaginary time evolution to obtain states that have high overlap with low energy states.  

In this paper, we implement the SSVQE framework for qumode platforms, ensuring orthonormality of the variational states by applying the same unitary ansatz $U (\bm{\theta})$ to a set of orthonormal states, as follows:
\begin{subequations} 
\begin{align}
\ket{\psi_n (\bm{\theta})} 
&= U (\bm{\theta}) \ket{\phi_n},
\\
\braket{ \psi_n (\bm{\theta}) | \psi_m (\bm{\theta}) }
&= \delta_{n m}, \quad \forall \: n, m,
\end{align}
\end{subequations}

We optimize the cost function $F$ defined by the sum of energy expectation values of $N_S$ orthonormal states, as follows:
\begin{subequations} \label{eq: subspace_vqe_qubit}
\begin{align}
\min_{\bm{\theta}} F 
&= \sum_{n = 1}^{N_S} \: w_n \: E_n (\bm{\theta}),
\\
E_n (\bm{\theta})
&= \braket{ \phi_n | U^\dagger (\bm{\theta}) \: 
\HamQ \: U (\bm{\theta}) | \phi_n },
\end{align}
\end{subequations}
where $\{ w_n \}$ are chosen weights, and $\{ \ket{\phi_n} \}$ is a set of easily prepared multi-qubit orthonormal states, e.g., 
$ \{ \ket{0}, \ket{1}, \ket{2}, \ket{3} \}$ with integer indexing.
The variational principle still holds for \Eq{\ref{eq: subspace_vqe_qubit}} because $F$ cannot be smaller than the combined energies for the corresponding exact stationary states. 
However, in some cases, certain individual energies $E_n$ may become smaller than their optimized energies, albeit with small error for a robust trial ansatz. 

\subsection{Bosonic quantum computing}

Quantum devices based on bosonic qumodes represent a distinct paradigm in quantum computing. Qumodes correspond to quantum harmonic oscillators and can be realized in various hardware platforms, including superconducting microwave cavities, \cite{Copetudo2024} 
trapped ions, \cite{Araz2025hybrid}
photonic systems, \cite{Lenzini2018integrated}
and mechanical resonators. \cite{Sarma2021continuous}
Unlike qubits, which are limited to two discrete energy levels, qumodes have an infinite-dimensional Hilbert space spanned by the Fock basis $\{ \ket{n} \}$, where $\ket{n}$ denotes the $n$-photon (or phonon) number eigenstate. 
In practice, computations are restricted to a finite subset of these levels, defined by the Fock cutoff.

Qumodes with a truncated subset can encode discrete quantum information similarly to qudits. For instance, the two-qubit computational basis 
$\{ \ket{0,0}_Q, \ket{0,1}_Q, \ket{1,0}_Q, \ket{1,1}_Q \}$ 
can be mapped to the Fock states 
$\{ \ket{0}_B, \ket{1}_B, \ket{2}_B, \ket{3}_B \}$ 
of a single bosonic mode with a Fock cutoff of four.
Measurement in the Fock basis---achieved, for example, through photon-number-resolving detectors---yields the probabilities $|\braket{n|\psi}|^2$,~\cite{Wang2020vibronic} analogous to Pauli-$Z$ measurements in qubit-based platforms.

One key advantage of qumode-based devices is their ability to efficiently implement native unitary operations that would require deep and resource-intensive circuits on qubit-based platforms.~\cite{Wang2020vibronic,Liu2024qumodequbitreview} 
A canonical example is the phase-space displacement operator,
\begin{equation} \label{eq: disp_op}
D(\alpha) = \exp\left( \alpha  \BC{} - \alpha^*  \BA{} \right),
\end{equation}
where $\BC{}$ and $\BA{}$ are the bosonic creation and annihilation operators, respectively. This operator belongs to the class of Gaussian unitaries, which are generated by quadratic bosonic Hamiltonians.

Universal bosonic quantum computation, however, requires access to non-Gaussian operations, which pose significant challenges for many physical platforms. A major advance in this direction has been achieved in circuit quantum electrodynamics (cQED), where a superconducting microwave cavity (qumode) is coupled to a transmon qubit. This hybrid architecture enables the implementation of the nonlinear, non-Gaussian interactions required for universal quantum control.~\cite{Copetudo2024,Liu2024qumodequbitreview}


\section{QSS-VQE Method} \label{sec: subspace_vqe_qumode}

In this section, we describe how qumode-based gates can be employed to implement the qumode subspace variational quantum eigensolver (QSS-VQE). In particular, we outline how to evaluate expectation values of a many-qubit Hamiltonian with respect to a qumode-based variational ansatz.


\begin{figure*}[t!]

\includegraphics[width=0.9\textwidth]{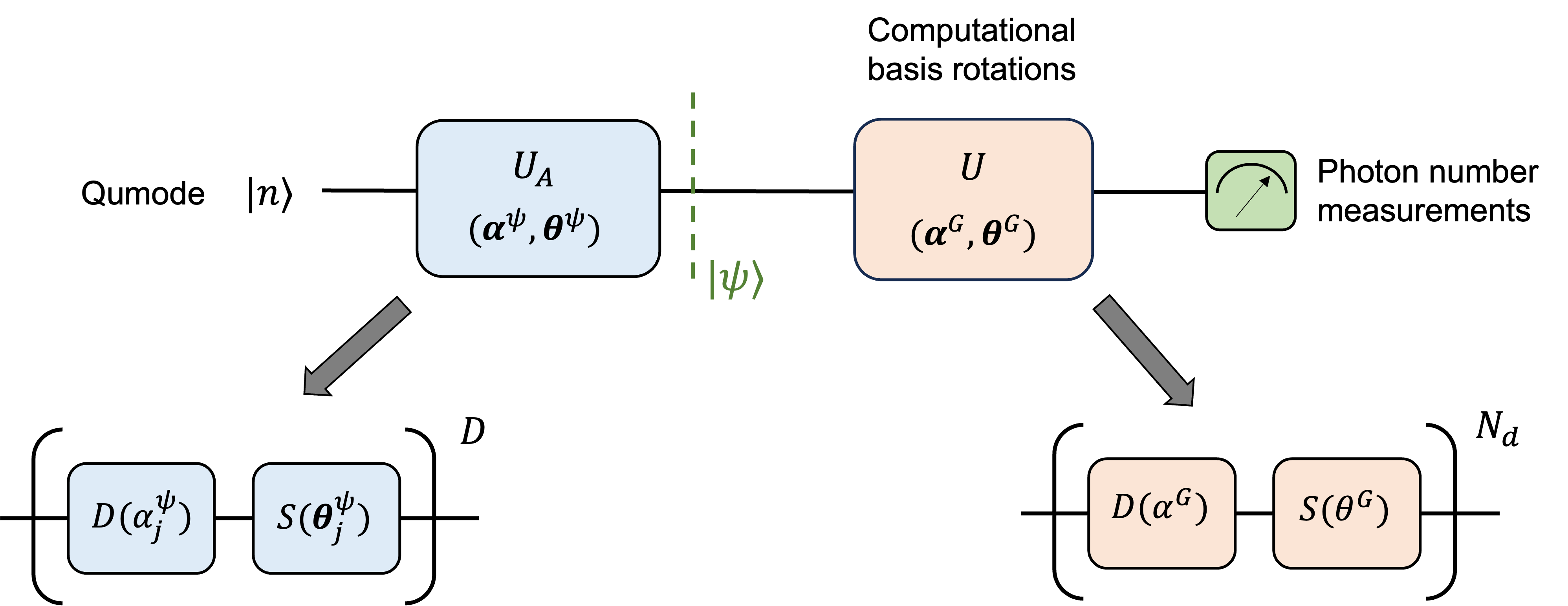}

\caption{
    Full circuit for computing the expectation value of a Pauli word with respect to a qumode state. 
    After state preparation of Fock basis state $\ket{n}$, the expectation value is computed by applying a set of SNAP-displacement gates followed by photon number measurements as discussed in \Sec{\ref{sec: exp_val}}.
}
\label{fig: full_circuit_snap}
\end{figure*}


\subsection{Universal Bosonic Ansatz} \label{sec: universal_ansatz} 

We implement the QSS-VQE approach on qumode platforms by mapping the Hamiltonian expectation values, originally expressed in a multi-qubit basis, onto the Fock basis of a single bosonic mode.
The variational unitary ansatz $U(\bm{\theta})$ is realized using the native gate set of a qubit–qumode architecture,\cite{Eickbusch2022,Kudra2022,Krastanov2015,Hillmann2020,Fosel2020,Huang2025fast,Liu2024qumodequbitreview}
enabling greater flexibility and expressiveness in ansatz design.
As we discuss below, evaluating the energy expectation values according Eq.~\ref{eq: subspace_vqe_qubit} requires photon-number-resolving measurements on the qubit–qumode system.\cite{Wang2020vibronic}

Preparing the trial state and evaluating energy expectation values within this framework requires access to a universal set of bosonic gates, as discussed below. A central component of this set is the selective number-dependent arbitrary phase (SNAP) gate,~\cite{Krastanov2015,Fosel2020} defined as
\begin{equation} \label{eq: snap_op}
S(\bm{\theta}) = \exp\left(i \sum_{n=0}^{L-1} \theta_n \ket{n}\bra{n} \right) = \sum_{n=0}^{L-1} e^{i\theta_n} \ket{n}\bra{n},
\end{equation}
where $\ket{n}\bra{n}$ is the projector onto the $n$-th Fock state, and $L$ is the Fock cutoff for the qumode.
Universal control of a single qumode can be achieved by combining SNAP gates with displacement operations $D(\alpha)$. Repeating layers of the form
\begin{equation} \label{eq: snap_disp_op}
\mathcal{U}_{\text{SD}}(\alpha, \bm{\theta}) = S(\bm{\theta}) D(\alpha),
\end{equation}
is sufficient to approximate arbitrary unitaries.~\cite{Krastanov2015,Job2023efficient} 
Without loss of generality, the displacement coefficients $\alpha$ can be taken as real, since complex phases can be absorbed into the SNAP parameters.
In this work we restrict ourselves to single-qumode circuits, so at most \(N_Q\) qubits can be replaced by the qumode, with its Fock cutoff \(L = 2^{N_Q}\). 
Current experiments support around the dimension of \(N_Q \approx 4\)~\cite{Wang2020vibronic,Curtis2021}, and recent progress suggests this could reach \(N_Q \approx 7\) in the near future~\cite{Deng2024quantum}. 
A separate route toward larger effective Hilbert spaces is to use multi-qumode ansatz circuits, which have been proposed for qumode-based VQE targeting ground states~\cite{Dutta2024EST}, which lies beyond the scope of this work. 

\subsection{Expectation Values} \label{sec: exp_val}

We focus on estimating the energy expectation value of the qumode state $\ket{\psi}$ under the qubit Hamiltonian $H_Q$ defined in Eq.~\eqref{eq: qubit_ham}:
\begin{equation} \label{eq: exp_val}
\braket{\psi| \: H_Q \: |\psi}
= \sum_{j = 1}^{N_H} \: \HQCoeff{j} 
\braket{ \psi | \: \Pauli{N_Q}{j} \: | \psi }, 
\end{equation}
where each $\Pauli{N_Q}{j}$ is a Pauli word acting on $N_Q$ qubits, and $g_j \in \mathbb{R}$ are the corresponding coefficients. Therefore, it suffices to evaluate the expectation values of individual Pauli words $\Pauli{N_Q}{j}$ on the quantum device, with the final energy computed via classical post-processing.
We now present a protocol for estimating arbitrary Pauli word expectation values using photon number measurements on a qumode-based quantum device. This extends our previous work on the integer-to-binary mapping approach.~\cite{Dutta2025solving}

As described above, the basis states of an $N_Q$-qubit system can be mapped to the Fock basis states of a qumode with cutoff dimension $L = 2^{N_Q}$ via a binary-to-integer encoding:
\begin{equation} \label{eq: binary_map}
\ket{q_1}_Q \otimes \cdots \otimes \ket{q_{N_Q}}_Q 
\leftrightarrow \ket{n}_B,
\end{equation}
where the Fock index is given by
$n = 2^{N_Q - 1} q_1 + \cdots + 2^0 q_{N_Q}$.
Now consider the case where the Pauli word $\Pauli{N_Q}{}$ consists only of identity and $N (\leq N_Q)$ Pauli-$Z$ operators. 
In this case, the expectation value can be written as
\begin{equation} \label{eq: exp_val_diag_pauli}
\braket{\psi| Z_{q_1} \cdots Z_{q_N} | \psi}
= \sum_{\mathbf{b}} (-1)^{\sum_{i=1}^N b_{q_i}}  \: P(\mathbf{b}),
\end{equation}
where $\mathbf{b} = (q_1, \ldots, q_{N_Q})$ denotes a bitstring corresponding to a computational basis state, and $P(\mathbf{b})$ is the probability of observing $\mathbf{b}$ in a measurement of $\ket{\psi}$.
Since each bitstring $\mathbf{b}$ is uniquely mapped to a Fock basis state $\ket{n}_B$ through Eq.~\eqref{eq: binary_map}, the full distribution ${P(\mathbf{b})}$ can be obtained directly via photon number measurements on the qumode state $\ket{\psi}$. This enables efficient evaluation of expectation values for Pauli words composed solely of $Z$ and identity operators.

Let us now extend our protocol for cases where some of the operators in the Pauli word $\Pauli{N_Q}{}$ are either Pauli-$X$ or Pauli-$Y$. 
It is known that computing the expectation values for these cases involves rotating them into the computational basis and absorbing the resulting unitary one-qubit operators into the trial state using
\begin{subequations}
\begin{align}
X_j 
&= H_j \: Z_j \: H_j, 
\\
Y_j 
&= ( H_j S_j^\dagger )^\dagger \: 
Z_j \: 
( H_j S_j^\dagger ), 
\end{align}
\end{subequations} 
where $H_j$ and $S_j$ are the Hadamard and phase gates, with the following one-qubit representations: 
\begin{equation}
H 
= \frac{1}{\sqrt{2}} \begin{bmatrix}
1 & 1 \\ 1 & - 1
\end{bmatrix}, \quad
S
= \begin{bmatrix}
1 & 0 \\ 0 & i
\end{bmatrix}.
\end{equation}
Adapting this qubit-centric protocol to the qumode case requires mapping the operators 
$H_j$ and $W_j = H_j S_j^\dagger$ into a sequence of qumode gates, as discussed below. 

We approximate an arbitrary target unitary operator \( V \) using a SNAP-displacement ansatz of the form  
\begin{equation} \label{eq: snap_disp_ansatz}
\UniGate{N_d}{} ( \bm{\alpha}^G, \bm{\theta}^G ) 
= \UniSNAP (\alpha_{N_d}, \bm{\theta}_{N_d}) \cdots 
\UniSNAP (\alpha_1, \bm{\theta}_1),
\end{equation}
where $N_d$ denotes the circuit depth. The variational parameters \( \bm{\alpha}^G \) and \( \bm{\theta}^G \) are optimized classically by minimizing the cost function~\cite{Dutta2024EST}
\begin{subequations} \label{eq: opt_cost_fun_snap}
\begin{align}
\min_{ \bm{\alpha}^G, \bm{\theta}^G } C 
&= \frac{1}{L^2} \sum_{n, m = 0}^{L - 1} 
| V_{n, m} - U_{n, m} |^2, \\
V_{n, m} 
&= \braket{n | V | m }, \\
U_{n, m}
&= \braket{n | \UniGate{N_d}{}  
( \bm{\alpha}^G, \bm{\theta}^G ) | m },
\end{align}
\end{subequations}
where \( \{ \ket{n} \} \) are Fock basis states and \( L \) is the Fock cutoff determined by the dimension of \( V \).
Once the optimized ansatz parameters are precomputed for all relevant \( \{ H_j, W_j \} \) corresponding to a fixed number of qubits, we can efficiently evaluate the expectation value of any Pauli word \( \Pauli{N_Q}{} \). 
This is done by applying the appropriate sequence of SNAP-displacement gates---chosen according to the number of \( X \) and \( Y \) operators in the Pauli word---prior to performing photon number measurements.
We note that we empirically observe a circuit depth scaling of \(N_d \sim L\) for implementing arbitrary target unitaries~\cite{Dutta2024EST}. For a practical single-qumode cutoff \(L \le 16\), as discussed in \Sec{\ref{sec: universal_ansatz}}, the circuits representing \(\{H_j, W_j\}\) therefore contribute only a fixed depth, independent of the size of the overall system. In larger settings, a partitioning strategy can be used to decompose the problem into independent optimizations over multiple qumodes. \cite{Dutta2024EST}
Consequently, this approach enables the estimation of the expectation value of a qubit Hamiltonian \( H_Q \), as defined in Eq.~\eqref{eq: exp_val}, using a single qumode device.

The qumode trial states $ \{ \ket{\psi_n} \} $ for each of the stationary states are also parameterized with a SNAP-displacement ansatz
\begin{subequations} \label{eq: snap_disp_ansatz_subspace_vqe}
\begin{align}
\ket{ \psi_n ( \bm{\alpha}^\psi, \bm{\theta}^\psi ) }
&= \UniAn{D} ( \bm{\alpha}^\psi, \bm{\theta}^\psi ) \ket{n}
\\
\UniAn{D} ( \bm{\alpha}^\psi, \bm{\theta}^\psi )
&= \UniSNAP (\alpha_{D}^\psi, \bm{\theta}_{D}^\psi) \cdots 
\UniSNAP (\alpha_1^\psi, \bm{\theta}_1^\psi), 
\end{align}
\end{subequations}
where $D$ is the circuit depth for the  trial ansatz, and $\UniSNAP$ is defined by \Eq{\ref{eq: snap_disp_op}}. 
The full circuit is illustrated in \Fig{\ref{fig: full_circuit_snap}}.
The variational parameters $\{ \bm{\alpha}^\psi, \bm{\theta}^\psi \}$
are updated during the minimization of the cost function $F$ defined as 
\begin{subequations} \label{eq: subspace_vqe_qumode}
\begin{align}
\min_{ \bm{\alpha}^\psi, \bm{\theta}^\psi } F 
&= \sum_{n = 1}^{N_S} \: w_n \: E_n (\bm{\alpha}^\psi, \bm{\theta}^\psi),
\\
E_n
&= \braket{ \psi_n (\bm{\alpha}^\psi, \bm{\theta}^\psi) | \HamQ | \psi_n (\bm{\alpha}^\psi, \bm{\theta}^\psi) },
\end{align}
\end{subequations}
where $\{ w_n \}$ are chosen weights and $\{ \ket{n} \}$ are the Fock basis states.
We employ numerical gradients of the cost function $F$ in some of our calculations, as shown in \Sec{\ref{sec: applications}}.
While parameter-shift rules can be constructed for bosonic Gaussian operations in close analogy to qubit gates~\cite{Mitarai2018}, extending them to non-Gaussian unitaries is more involved~\cite{Schuld2019}. 
Efficient gradient-based optimization for general qumode unitaries can nevertheless be achieved using automatic differentiation~\cite{Eickbusch2022}.

Quantum state preparation of a qumode Fock basis state $\ket{n}$ from the vacuum state $\ket{0}$ can be accomplished using a variety of techniques within the circuit quantum electrodynamics (cQED) framework.~\cite{Ma2021quantum,Eickbusch2022,Krastanov2015,Job2023efficient,Zhang2024generating,Huang2025fast}
A standard approach employs pulse-shaping methods such as Gradient Ascent Pulse Engineering (GRAPE), which utilizes numerical optimization to synthesize the desired state.~\cite{Ma2021quantum} 
More recently, echoed conditional displacement (ECD) gates combined with qubit rotation ansatze have demonstrated favorable scaling in circuit depth for Fock state preparation when used in conjunction with numerical optimization methods,~\cite{Eickbusch2022} 
which can be similarly implemented with SNAP-displacement ansatz as well, either via numerical optimization~\cite{Krastanov2015} or through an analytical decomposition based on Givens rotations.~\cite{Job2023efficient} 
Fock states can also be generated probabilistically from coherent states by performing projective measurements on an ancilla qubit coupled to the cavity mode.~\cite{Zhang2024generating} 
Maybe the most efficient known method to date leverages an analytical scheme based on sideband transitions combined with qubit rotations.~\cite{Huang2025fast} This technique offers a circuit depth linear in $n$ with known rotation angles, thus eliminating the need for explicit optimization.~\cite{LawEberly1996,Mischuck2013,Liu2021qudit}


\section{Applications} \label{sec: applications}

We present benchmark calculations comparing the performance of QSS-VQE with qubit-based approaches, alongside applications to model Hamiltonians that demonstrate the effectiveness of the qumode-based method for computing ground and excited states of molecular systems.

\subsection{Dihydrogen molecule}

\begin{figure}[b!]

\includegraphics[width=0.9\columnwidth]{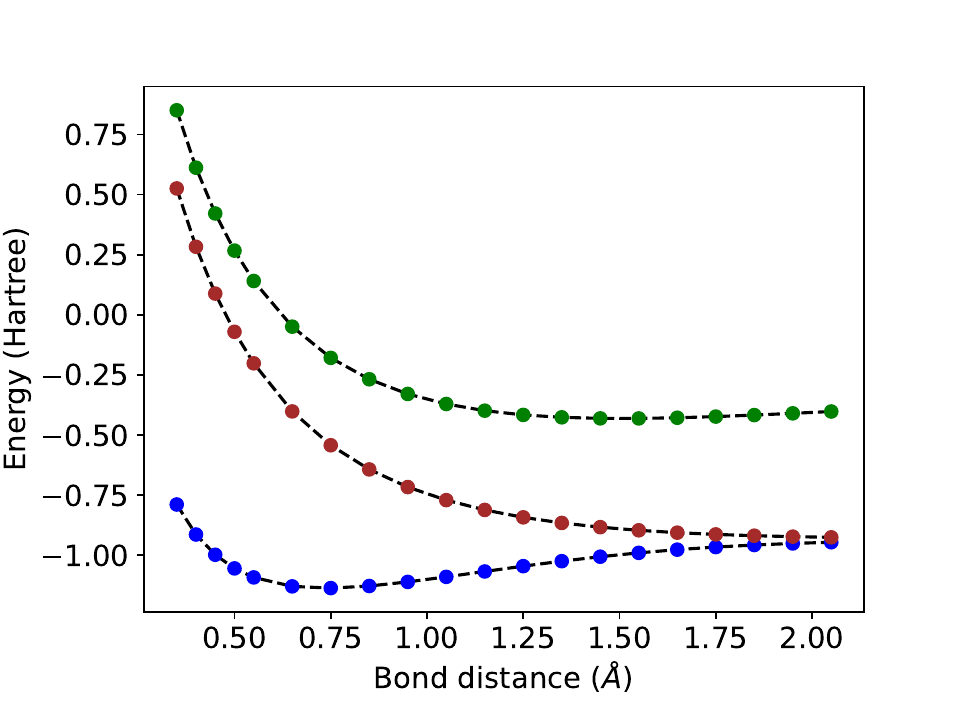}

\caption{
    Comparison between the energies from exact diagonalization (dashed lines) and subspace VQE (circles) as defined in \Sec{\ref{sec: subspace_vqe_qumode}} for the ground and first two excited states of the dihydrogen molecule in STO-3G basis. 
    The circuit depth for the subspace VQE SNAP-displacement ansatz is $D = 4$ with the weight parameters $\textbf{w} = ( 1.0, 0.9, 0.8 )$.
}
\label{fig: h2mol_ssvqe_en_snap_nd4}
\end{figure}


\begin{figure}[t!]

\includegraphics[width=0.9\columnwidth]{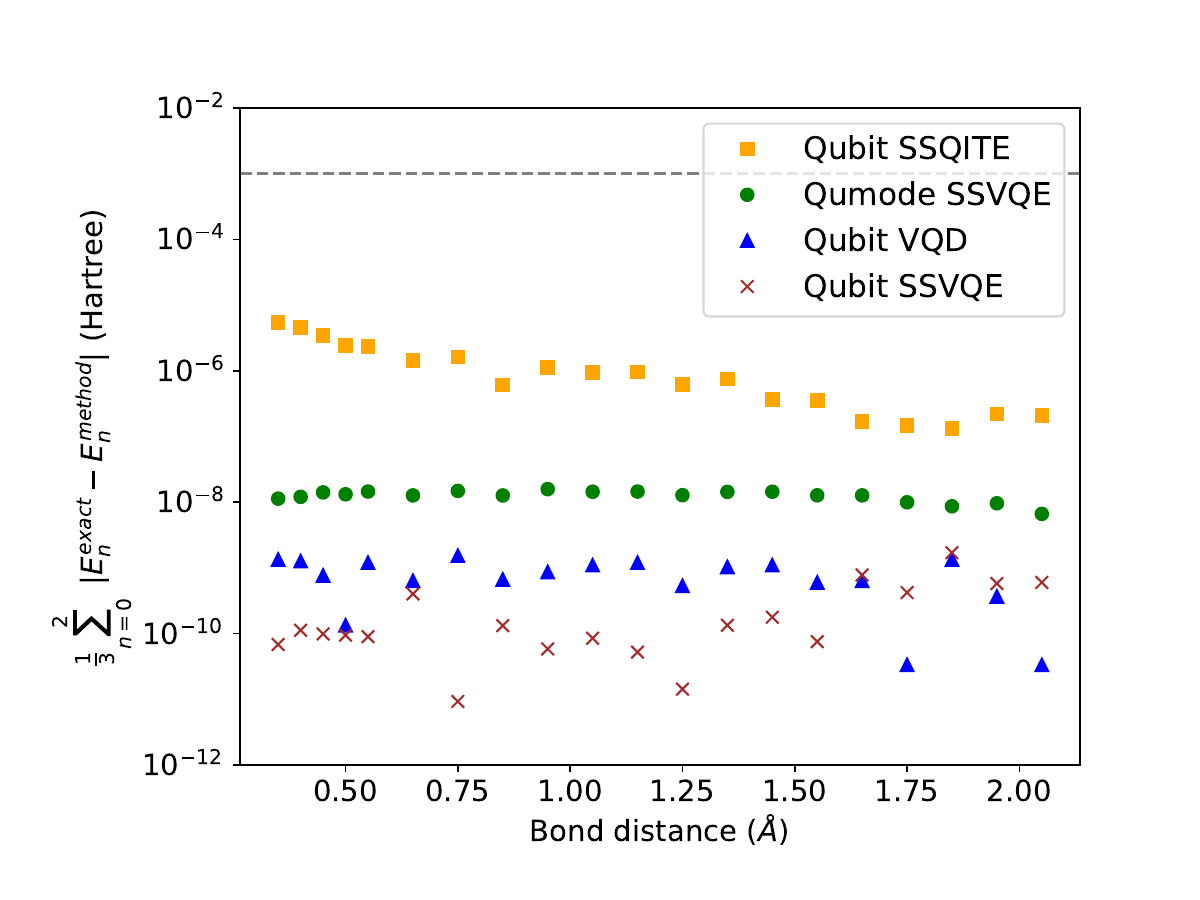}

\caption{
    Comparison between different qubit-based excited state methods and the subspace VQE method as defined in \Sec{\ref{sec: subspace_vqe_qumode}} for the ground and first two excited states of dihydrogen molecules in STO-3G basis. 
    The vertical axis plots the average error function defined in \Eq{\ref{eq: error_sum_function}}.
    The circuit depth for the subspace VQE with SNAP-displacement ansatz for the trial states is $D = 4$ with the weight parameters $\textbf{w} = ( 1.0, 0.9, 0.8 )$.
}
\label{fig: h2mol_sum_en_error}
\end{figure}


The electronic Hamiltonian of the dihydrogen molecule in the STO-3G basis set can be mapped to two-qubit states by projecting the problem into the spin-zero subspace, as follows: 
\begin{equation} \label{eq: h2_qubit_ham}
\HamQ
= \HQCoeff{0} (R)
+ \HQCoeff{1} (R) \: Z_2
+ \HQCoeff{2} (R) \: X_1 X_2 
+ \HQCoeff{3} (R) \: Z_1 
+ \HQCoeff{4} (R) \: Z_1 Z_2, 
\end{equation}
where $\{ \HQCoeff{j} (R) \}$ are scalar coefficients that depend on the H--H bond distance $R$. \cite{Colless2018spectra,cianci2024subspacesearchquantumimaginarytime} 
SNAP-displacement ansatze are parametrized to represent the Pauli words introduced by Eq.~(\ref{eq: h2_qubit_ham}), including $Z \otimes \EYE $, $\EYE \otimes Z$, $Z \otimes Z$, and $X \otimes X$. 
Using $N_d = 4$, we achieve losses of the order of $ 1 \times 10^{-13} $, or smaller.  
With these ansatze, we have applied the QSS-VQE method, described in \Sec{\ref{sec: subspace_vqe_qumode}}, to obtain the lower three stationary states of the dihydrogen molecule.

As shown in \Fig{\ref{fig: h2mol_ssvqe_en_snap_nd4}}, energies from the subspace VQE reproduce the exact energies found by diagonalization for the ground and first two excited states of $\text{H}_2$ at different H--H bond distances. 
The circuit depth for the ansatz is $D = 4$ and the weight parameters for the cost function are 
$\textbf{w} = ( 1.0, 0.9, 0.8 )$.
The number of parameters was 20, which were optimized using the BFGS algorithm in SciPy. \cite{2020SciPy-NMeth}
We evaluated the performance of QSS-VQE relative to qubit-based approaches by comparing energy errors as a function of the H--H bond distance, including benchmarks against SSQITE, SSVQE, and VQD (see Fig.~\ref{fig: h2mol_sum_en_error}). The comparison is based on the average absolute energy error $\Delta$ across the three lowest stationary states, defined as:
\begin{equation} \label{eq: error_sum_function}
\Delta 
= \frac{1}{3} \sum_{n = 0}^2 \left| E_n^{\text{exact}} - E_n^{\text{method}} \right|.
\end{equation}
The qubit-based subspace VQE and VQD algorithms employed a BFGS optimizer within a noiseless simulator. In contrast, the subspace QITE algorithm, which does not use a classical optimizer, relies on McLachlan’s variational principle to evolve the system and minimize the energy. All qubit-based methods utilized a two-layer TwoLocal ansatz incorporating tunable $R_x$ and $R_y$ single-qubit rotation gates, 
which means a total of four parameters were optimized.
All of the iterations were chosen to be up to 100, considering the practical near-term quantum computing resources.
As shown in Fig.~\ref{fig: h2mol_sum_en_error}, 
the qumode-based QSS-VQE, along with the qubit-based subspace VQE and VQD methods, achieved lower energy errors than the qubit-based subspace QITE. All methods reached chemical accuracy across the range of bond distances considered.

\subsection{Conical Intersection in Cytosine}


\begin{figure}[t!]
\includegraphics[width=0.9\columnwidth]{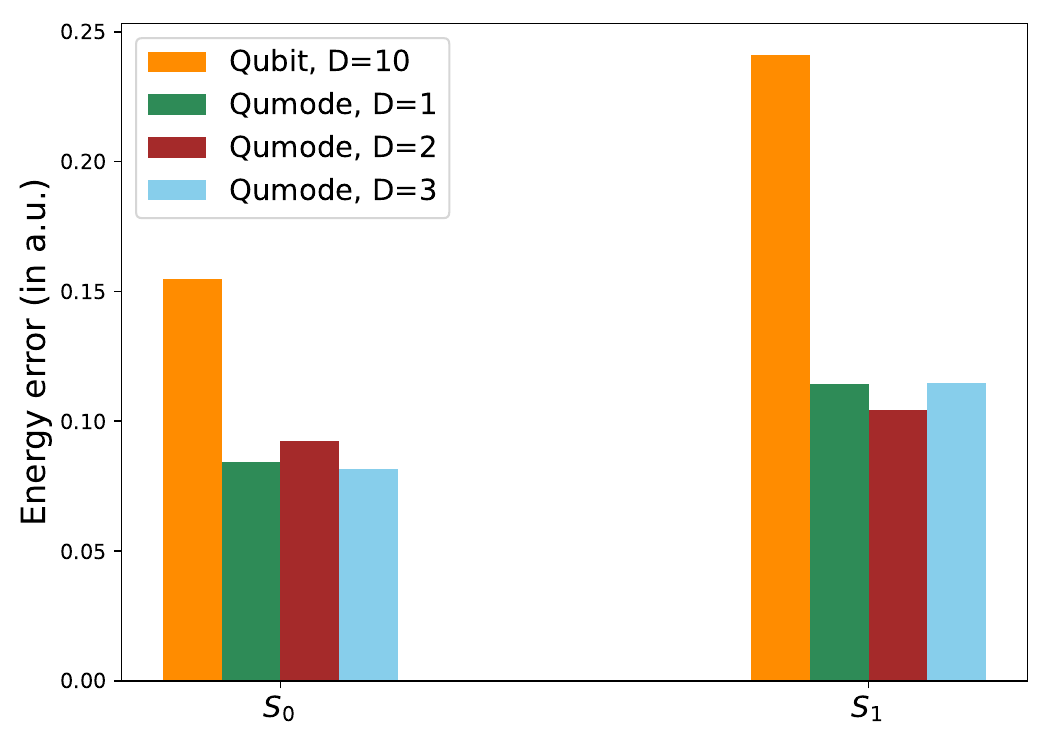}
\caption{
    Comparison of exact energies with those obtained from qubit-based and qumode-based SSVQE methods. The qubit-based subspace VQE uses a TwoLocal ansatz with circuit depth $D = 10$, while the qumode-based QSS-VQE employs a SNAP-displacement ansatz with circuit depths up to $D = 3$.
    Both methods use uniform weight parameters $\mathbf{w} = (1.0, 1.0, 1.0)$ in the cost function.
}
\label{fig: cyt_en_compare}
\end{figure}


Here, we focus on the calculation of excited states at a conical intersection between potential energy surfaces of cytosine, as recently reported in \Reference{\citenum{Wang2025characterizing}}. 
The corresponding Hamiltonian, which yields degenerate ground and excited singlet-state energies, was constructed using a complete active space self-consistent field (CASSCF) method with an active space of four electrons in three orbitals, employing the cc-pVDZ basis set.
To reduce the qubit requirements---originally six qubits under the Jordan-Wigner transformation---two symmetry constraints were applied: conservation of the total number of electrons and the total spin. 
This reduced the effective number of qubits to four.
The resulting four-qubit Hamiltonian contains $N_H = 136$ Pauli terms, including identity operators on all qubits. \cite{Wang2025characterizing}

The SNAP-displacement parameters for all $\{ H_j, W_j \}$ operator pairs (with $N_Q = 4$ qubits) were optimized using the cost function defined in Eq.~(\ref{eq: opt_cost_fun_snap}), with a circuit depth of $N_d = 16$. The final optimized losses for these Pauli terms are summarized in Table~\ref{tab: cyt_losses}.


\begin{table}[h!]
\begin{tabular}{ll|ll}
\hline
Operator &  Loss &  Operator &  Loss                   
\\ \hline
$H_1$ &  $4 \times 10^{-14}$ &  $W_1$ &  $2 \times 10^{-14}$                  
\\
$H_2$ &  $9 \times 10^{-14}$ &  $W_2$ &  $1 \times 10^{-14}$               
\\
$H_3$ &  $6 \times 10^{-14}$ &  $W_3$ &  $2 \times 10^{-13}$                   
\\
$H_4$ &  $1 \times 10^{-13}$ &  $W_4$ &  $5 \times 10^{-14}$                    
\\
\hline
\end{tabular}

\caption{
    Converged loss values for representing the four-qubit operators with the SNAP-displacement ansatz of depth $N_d = 16$. 
}
\label{tab: cyt_losses}
\end{table}


Resolving the spectrum of this Hamiltonian presents a significant challenge for excited-state methods, as the two singlet excited states are nearly degenerate and contribute to the formation of the conical intersection~\cite{Wang2025characterizing}. 
In Fig.~\ref{fig: cyt_en_compare}, we show the energy errors of the QSS-VQE method with SNAP-displacement ansatz for the three lowest stationary states of cytosine, using uniform cost function coefficients $\mathbf{w} = (1.0, 1.0, 1.0)$. 
As illustrated in Fig.~\ref{fig: cyt_en_compare}, the SNAP-displacement gates enable a more efficient representation of the excited states compared to the qubit-based ansatz.
For the single-qumode case (\(D = 1\)), the ansatz contains 17 parameters, compared with 20 parameters in the \(D = 10\) qubit-based TwoLocal ansatz. Both were optimized using COBYLA and converged within 200 iterations. The achievable accuracy does not improve with additional depth in either setup because of local minima. This limitation can be removed by switching to a more brute-force gradient-based method such as BFGS, which yields comparable performance for both the ansatz circuits given the small system size.

\subsection{Model Hamiltonians} \label{sec: model_qubit_hams}

The SNAP gates used in the universal qumode ansatz carry multiple tunable parameters, unlike typical qubit-based parametrized gates, which can give the qumode approach a practical advantage. 
Here we show that even with a single parametrized gate, a single qumode unitary can represent stationary states of a class of qubit Hamiltonians more efficiently than a universal qubit unitary block.  
The Hamiltonian corresponds to a phase-space displaced harmonic oscillator:
\begin{equation} \label{eq: disp_qho_ham}
H_B 
= \frac{1}{2} \: \BP{}^2
+ \frac{1}{2} \: ( \BX{} - \sqrt{2} \: \alpha )^2,
\end{equation}
where $\BX{}$ and $\BP{}$ denote the position and momentum operators of the harmonic oscillator, respectively, and $\alpha$ is a real-valued displacement parameter. In atomic units, these quadrature operators are expressed in terms of the bosonic creation and annihilation operators
\begin{equation} \label{eq: bosonic_quad_ops}
\BX{} 
= \frac{1}{\sqrt{2}} \: ( \BC{} + \BA{} ), \quad
\BP{} 
= \frac{i}{\sqrt{2}} \: ( \BC{} - \BA{} ).
\end{equation}
The exact ground and excited states of the Hamiltonian, introduced by Eq.~(\ref{eq: disp_qho_ham}), can be represented as
\begin{equation} \label{eq: disp_state}
\ket{\psi_n}
= D(\alpha) \ket{n},
\end{equation}
where $\{ \ket{n} \}$ are the Fock states and $D(\alpha)$ is the displacement operator, defined by Eq.~(\ref{eq: disp_op}).
This can be physically motivated by the fact that Eq.~(\ref{eq: disp_qho_ham}) is a displaced oscillator, which means all of the original stationary states are also displaced by the same amount.

To compute the ground and excited state energies of this model Hamiltonian with a trial ansatz, we need to first write $H_B$ in terms of Pauli words. 
To that end, we first write the creation and annihilation operators in a truncated basis of Fock states as 
\begin{subequations} \label{eq: bosonic_ops_proj}
\begin{align}
\BC{}
&= \sum_{n = 0}^{L - 2} \sqrt{n + 1} 
\KetBra{n + 1}{n},
\\
\BA{}
&= \sum_{n = 1}^{L - 1} \sqrt{n} 
\KetBra{n - 1}{n},
\end{align}
\end{subequations}
where $L$ is the Fock truncation cutoff. Next, we map the projection operators $\KetBra{n}{m}$ into a tensor product of one-qubit projection operators:~\cite{Wang2023quantum,Peng2025quantum}
\begin{subequations} \label{eq: qubit_proj_pauli}
\begin{align}
\KetBra{0}{0}
&= \frac{1}{2} \: ( 1 + Z ), \quad  
\KetBra{1}{1}
= \frac{1}{2} \: ( 1 - Z ),
\\
\KetBra{0}{1}
&= \SP{}, \quad   
\KetBra{1}{0}
= \SM{},
\end{align}
\end{subequations} 
where 
$ \sigma^{\pm} 
= \frac{1}{2} \: ( X \pm i Y ) $.
This allows us to map $H_B$ to an $N_Q$-qubit Hamiltonian $H_Q$, for a chosen truncation cutoff $L=2^{N_Q}$. 

Let us first obtain $H_Q$ for $N_Q = 2$.      
The four Fock basis states can be mapped to two-qubit basis states, as follows: 
\begin{subequations}
\begin{align}
\ket{0}_B 
&= \ket{0}_Q \otimes \ket{0}_Q, \quad   
\ket{1}_B 
= \ket{0}_Q \otimes \ket{1}_Q,    
\\
\ket{2}_B 
&= \ket{1}_Q \otimes \ket{0}_Q, \quad   
\ket{3}_B 
= \ket{1}_Q \otimes \ket{1}_Q,
\end{align}
\end{subequations}
which combined with \Eq{\ref{eq: bosonic_ops_proj}} and \Eq{\ref{eq: qubit_proj_pauli}} represent the bosonic creation operator as 
\begin{align}
\BC{}
&= \Big( \KetBra{1}{0}
+ \sqrt{2} \KetBra{2}{1}
+ \sqrt{3} \KetBra{3}{2} \Big)_B 
\nonumber 
\\ 
&\mapsto \frac{1}{2} \: (1 + Z_1) \: \SM{2}
+ \sqrt{2} \: \SM{1} \SP{2} 
+ \frac{\sqrt{3}}{2} \: (1 + Z_1) \: \SM{2},
\end{align}
giving a representation of the position and momentum operators, introduced by Eq.~(\ref{eq: bosonic_quad_ops}), using the expressions below 
\begin{subequations}
\begin{align}
\BC{} + \BA{} 
&\mapsto \frac{1}{2} \: \Big[ 
( 1 + \sqrt{3} ) \: X_2
+ \sqrt{2} \: ( X_1 X_2 + Y_1 Y_2 )
+ ( 1 - \sqrt{3} ) \: Z_1 X_2 
\Big],
\\ 
\BC{} - \BA{} 
&\mapsto - \frac{i}{2} \: \Big[ 
( 1 + \sqrt{3} ) \: Y_2
+ \sqrt{2} \: ( Y_1 X_2 - X_1 Y_2 ) 
+ ( 1 - \sqrt{3} ) \: Z_1 Y_2 
\Big].
\end{align}
\end{subequations}
Thus, when $N_Q = 2$, the bosonic Hamiltonian $H_B$ defined in \Eq{\ref{eq: disp_qho_ham}} can be represented as 
\begin{align} \label{eq: disp_qho_ham_mapped_q2}
H_B 
&\mapsto H_Q 
= \frac{1}{2} \: \Big[ 
( 2 \: \alpha^2 + 3 )
- Z_1 - Z_1 Z_2 
- ( 1 + \sqrt{3} ) \: \alpha \: X_2
- ( 1 - \sqrt{3} ) \: \alpha \: Z_1 X_2 
\nonumber
\\
&\hskip10ex 
- \sqrt{2} \: \alpha \: ( X_1 X_2 + Y_1 Y_2 )
\Big]. 
\end{align}
Similarly, we can derive the mapped expressions for the $N_Q = 3$. 
The three-qubit basis states can be mapped to eight Fock states as
\begin{subequations}
\begin{align}
\ket{0}_B 
&\leftrightarrow \ket{0, 0, 0}_Q, \quad
\ket{1}_B 
\leftrightarrow \ket{0, 0, 1}_Q, \quad
\ket{2}_B 
\leftrightarrow \ket{0, 1, 0}_Q, \quad
\ket{3}_B 
\leftrightarrow \ket{0, 1, 1}_Q, \quad
\\
\ket{4}_B 
&\leftrightarrow \ket{1, 0, 0}_Q, \quad
\ket{5}_B 
\leftrightarrow \ket{1, 0, 1}_Q, \quad 
\ket{6}_B 
\leftrightarrow \ket{1, 1, 0}_Q, \quad
\ket{7}_B 
\leftrightarrow \ket{1, 1, 1}_Q, 
\end{align}
\end{subequations}
where we have omitted the tensor product symbols for the qubit states.  
The bosonic creation operator can now be mapped, as follows: 
\begin{align}
\BC{} 
&= \Big( \KetBra{1}{0}
+ \sqrt{2} \KetBra{2}{1}
+ \sqrt{3} \KetBra{3}{2}
+ \sqrt{4} \KetBra{4}{3}
+ \sqrt{5} \KetBra{5}{4}
+ \sqrt{6} \KetBra{6}{5}
+ \sqrt{7} \KetBra{7}{6} \Big)_B
\nonumber
\\ 
&\mapsto \frac{1}{4} \: (1 + Z_1) \: 
(\EYE_2 + Z_2) \: \SM{3}    
+ \frac{1}{\sqrt{2}} \: (1 + Z_1) \: 
\SM{2} \: \SP{3} 
+ \frac{\sqrt{3}}{4} \: (1 + Z_1) \: 
(1 - Z_2) \: \SM{3} 
+ 2 \: \SM{1} \: \SP{2} \SP{3} 
\nonumber
\\
&+ \frac{\sqrt{5}}{4} \: (\EYE_1 - Z_1) \: 
(\EYE_2 + Z_2) \: \SM{3} 
+ \sqrt{ \frac{3}{2} } \: (\EYE_1 - Z_1) \: 
\SM{2} \: \SP{3} 
+ \frac{\sqrt{7}}{4} \: (\EYE_1 - Z_1) \: 
(\EYE_2 - Z_2) \: \SM{3}, 
\end{align}
which finally leads to the following expression below 
\begin{align}
\BC{} + \BA{} 
&\mapsto \frac{1}{4} \: \Big[
( 1 + \sqrt{3} + \sqrt{5} + \sqrt{7} ) \: X_3
+ ( \sqrt{2} + \sqrt{6} ) \: 
( X_2 X_3 + Y_2 Y_3 ) 
\nonumber
\\
&\hskip5ex
+ ( 1 + \sqrt{3} - \sqrt{5} - \sqrt{7} ) \: Z_1 X_3 
+ ( 1 - \sqrt{3} + \sqrt{5} - \sqrt{7} ) \: Z_2 X_3 
\nonumber
\\
&\hskip5ex
+ 2 \: ( X_1 X_2 X_3 - X_1 Y_2 Y_3 
+ Y_1 X_2 Y_3 + Y_1 Y_2 X_3 )
\nonumber
\\
&\hskip5ex
+ ( \sqrt{2} - \sqrt{6} ) \: 
( Z_1 X_2 X_3 + Z_1 Y_2 Y_3 ) 
+ ( 1 - \sqrt{3} - \sqrt{5} + \sqrt{7} ) \: Z_1 Z_2 X_3 \Big], 
\end{align}
with the corresponding qubit representation of $H_B$ in the truncated basis with $N_Q = 3$ as
\begin{align} \label{eq: disp_qho_ham_mapped_q3}
H_B 
&\mapsto H_Q 
= \frac{1}{4} \: \Big[
( 4 \: \alpha^2 + 14 )
- 2 \: ( 3 \: Z_1 + Z_2 
+ Z_1 Z_2 + Z_1 Z_3 
+ Z_2 Z_3 
- Z_1 Z_2 Z_3 ) 
\nonumber
\\
&\hskip7ex
- ( 1 + \sqrt{3} + \sqrt{5} + \sqrt{7} ) \: \alpha \: X_3
- ( \sqrt{2} + \sqrt{6} ) \: \alpha \:  
( X_2 X_3 + Y_2 Y_3 ) 
\nonumber
\\
&\hskip7ex
- ( 1 + \sqrt{3} - \sqrt{5} - \sqrt{7} ) \: \alpha \: Z_1 X_3 
- ( 1 - \sqrt{3} + \sqrt{5} - \sqrt{7} ) \: \alpha \: Z_2 X_3 
\nonumber
\\
&\hskip7ex
- 2 \: \alpha \: ( X_1 X_2 X_3 
- X_1 Y_2 Y_3 
+ Y_1 X_2 Y_3 
+ Y_1 Y_2 X_3 )
\nonumber
\\
&\hskip7ex
- ( \sqrt{2} - \sqrt{6} ) \: \alpha \: 
( Z_1 X_2 X_3 + Z_1 Y_2 Y_3 ) 
+ ( 1 - \sqrt{3} - \sqrt{5} + \sqrt{7} ) \: \alpha \: Z_1 Z_2 X_3 \Big]. 
\end{align}
Thus, the bosonic Hamiltonian $H_B$ of \Eq{\ref{eq: disp_qho_ham}} can be mapped to a class of $N_Q$-qubit Hamiltonians, defined according to \Eq{\ref{eq: disp_qho_ham_mapped_q2}} and \Eq{\ref{eq: disp_qho_ham_mapped_q3}}
for $N_Q = 2$ and $N_Q = 3$, respectively. 
The Hamiltonians with higher $N_Q$ can be derived analogously. 


\begin{figure}[t!]
    \centering
    \includegraphics[width=0.9\textwidth]{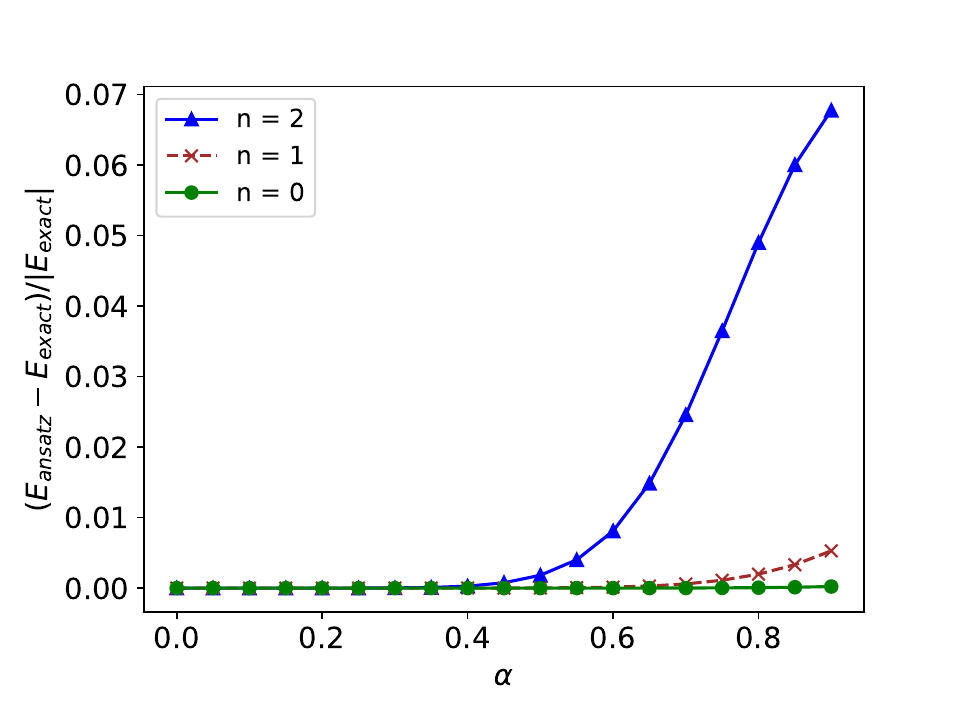}

    \caption{
       Relative errors of the trial state energies, $\braket{\psi_n | H_Q | \psi_n}$, compared to the exact eigenvalues of the mapped Hamiltonian $H_Q$ (defined in \Sec{\ref{sec: model_qubit_hams}}), as a function of $\alpha$. Trial states are defined as $\ket{\psi_n} = D(\alpha) \ket{n}$, where $D(\alpha)$ is a displacement operator and $\ket{n}$ denotes the $n$-th Fock state. Results are shown for the ground state and the first two excited states using $N_Q = 3$ qubits. 
    }

    \label{fig: disp_qho_en}
    
\end{figure}


\begin{figure*}[t!]
    \centering

    \includegraphics[width=0.97\textwidth]{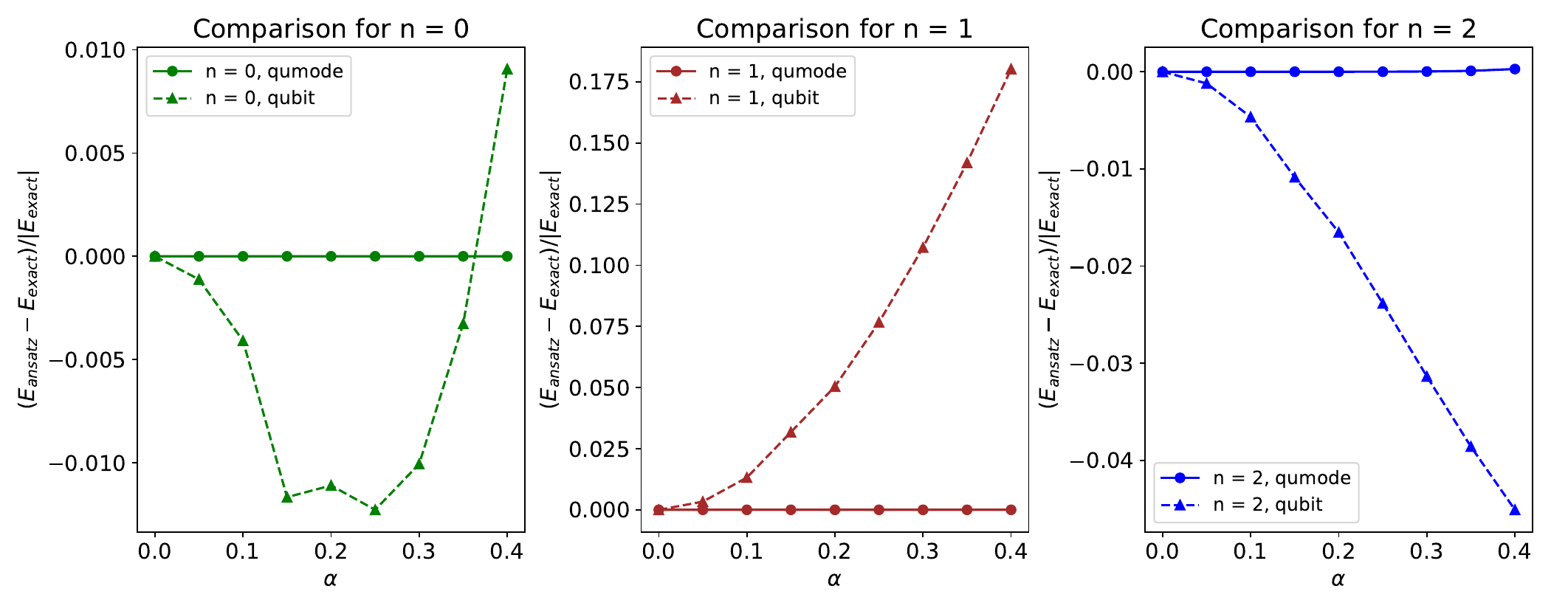}

    \caption{
        Comparison of exact energy errors for trial energies 
        $ \braket{\psi_n | \: H_Q \: | \psi_n} $ for the three-qubit Hamiltonian defined in \Eq{\ref{eq: disp_qho_ham_mapped_q3}}. 
        The $\ket{\psi_n}$ for the qumode case is one displacement operator gate acting on the Fock state $\ket{n}$. 
        The qubit method chosen is subspace VQE with a one-layer TwoLocal ansatz.
    }

    \label{fig: disp_qho_nq3_en_errors}
\end{figure*}


The mapped Hamiltonian $H_Q$, defined on a Fock basis of finite size $N_Q$, provides only an approximate representation of the original bosonic Hamiltonian $H_B$. As a result, the displaced states defined by \Eq{\ref{eq: disp_state}} are accurately represented by the ground and excited states of $H_Q$ only for sufficiently small values of the displacement parameter $\alpha$.
\Fig{\ref{fig: disp_qho_en}} illustrates the accuracy of this approximation by showing the deviation between the exact expectation values of $H_Q$ and the energies obtained with the variational states that are possible to implement within the QSS-VQE approach.
Clearly, \Fig{\ref{fig: disp_qho_en}} shows that $H_Q$, defined by \Eq{\ref{eq: disp_qho_ham_mapped_q3}} provides a very good approximation of the displaced ground ($n=0$) and excited-states  ($n = 1, 2$), when $\alpha \leq 0.4$. 

However, for a qubit-based implementation, this type of calculation may not be straightforward, given a generic universal ansatz with shallow depth. 
As shown in \Fig{\ref{fig: disp_qho_nq3_en_errors}}, the qubit-based SSVQE method based on a one-layer TwoLocal ansatz exhibits significant deviations, 
even with a gradient-based BFGS optimizer and up to 300 iterations of optimizations.
Thus, we show that there exists a class of multi-qubit Hamiltonians, for which a connection to oscillator algebra may be unknown, where trial ansatz circuits built from qumode unitaries can outperform qubit-based designs. 
This approach may also benefit inherently bosonic Hamiltonians that are classically difficult to treat, such as anharmonic vibrational systems~\cite{Barone2012anharmonic,Wang2022anharmonic}.


\section{Conclusions} \label{sec: final} 

We have introduced the QSS-VQE, or the Qumode Subspace Variational Quantum Eigensolver, that enables the calculation of ground and excited states of molecules, using a single microwave cavity coupled to an ancilla qubit---an architecture naturally suited to cQED devices. 
By leveraging the Hilbert space of the qumode and the native gate set based on the SNAP-displacement variational ansatz, we show the possibility of simulating multi-qubit Hamiltonians even when using the Fock states of a single qumode.

The QSS-VQE computes energy expectation values by measuring photon number distributions in the cavity, effectively replacing multi-qubit registers with a compact bosonic encoding. 
We benchmarked this method on the dihydrogen molecule in a minimal basis and found that the qumode-based ansatz achieved comparable or improved energy accuracy relative to qubit-only approaches.
On the more complex model system involving a conical intersection in cytosine, characterized by nearly degenerate singlet excited states, our method outperformed deeper qubit-only variational circuits, achieving better accuracy with fewer variational parameters. 
Finally, we have introduced a class of model bosonic Hamiltonians based on a displaced quantum harmonic oscillator for which a single qumode gate can outperform the corresponding qubit-based circuits in calculations of ground and excited states.

These results suggest that qumode-based variational quantum algorithms (VQAs), offer promising capabilities in reducing the circuit depth and parameter count for calculations of ground and excited states. Our study focused on encoding $N_Q$ qubits in a single qumode with Fock cutoff $L = 2^{N_Q}$, using the universal SNAP--displacement ansatz. 
The QSS-VQE framework can be extended to larger systems, following the scaling strategy used for ground states~\cite{Dutta2024EST}. Consider a partition of the qubits into \(k\) disjoint subsets of sizes \(N_{Q_1}, N_{Q_2}, \ldots, N_{Q_k}\). Each subset is mapped to a distinct qumode with Fock cutoff 
\((L_1, L_2, \ldots, L_k) = (2^{N_{Q_1}}, 2^{N_{Q_2}}, \ldots, 2^{N_{Q_k}})\), 
so that the total number of replaced qubits is \(N_Q = \sum_{j=1}^k N_{Q_j}\). This partitioning ensures that every qumode operates within a practical cutoff regime. 
In this extension, the expectation-value evaluation introduced in \Sec{\ref{sec: exp_val}} carries over directly: each Pauli operator acts on a specific qubit subset and hence maps to SNAP–displacement operations on the corresponding qumode only. Consequently, after optimizing \Eq{\ref{eq: opt_cost_fun_snap}} for each qumode independently, the global computational-basis rotation is the tensor product of the resulting single-qumode SNAP–displacement circuits, with no inter-qumode coupling required. 
In contrast, constructing the excited-state ansatz requires a universal gate set over the full \(k\)-qumode Hilbert space. 
Single-qumode SNAP–displacement gates are insufficient for universality in the multi-qumode setting. 
Universal multi-qumode constructions is currently an active research topic~\cite{Eickbusch2022,Liu2024qumodequbitreview,You2024Crosstalk,Dutta2024EST}, and can be obtained by combining entangling bosonic interactions (e.g., beamsplitters) with a universal single-qumode set such as the SNAP–displacement gates~\cite{Dutta2024EST,Liu2024qumodequbitreview}. 
This provides a systematic path toward scalable QSS-VQE implementations.

Looking ahead, the combination of QSS-VQE with universal multi-qumode ansatze opens exciting opportunities for efficient, hardware-native quantum algorithms. 
Our results thus mark an important step toward leveraging the full potential of bosonic quantum processors for simulating complex quantum systems beyond the capabilities of qubit-only architectures.


\begin{acknowledgement}

We acknowledge support from the NSF for the Center for Quantum Dynamics on Modular Quantum Devices (CQD-MQD) under grant number CHE-2124511. 

\end{acknowledgement}


\section*{Code and Data Availability}

The Python code and data for the plots can be found at \href{https://github.com/rishabdchem/qss_vqe_paper}{https://github.com/rishabdchem/qss\_vqe\_paper}.


\bibliography{QBM}

\providecommand{\latin}[1]{#1}
\makeatletter
\providecommand{\doi}
  {\begingroup\let\do\@makeother\dospecials
  \catcode`\{=1 \catcode`\}=2 \doi@aux}
\providecommand{\doi@aux}[1]{\endgroup\texttt{#1}}
\makeatother
\providecommand*\mcitethebibliography{\thebibliography}
\csname @ifundefined\endcsname{endmcitethebibliography}  {\let\endmcitethebibliography\endthebibliography}{}
\begin{mcitethebibliography}{63}
\providecommand*\natexlab[1]{#1}
\providecommand*\mciteSetBstSublistMode[1]{}
\providecommand*\mciteSetBstMaxWidthForm[2]{}
\providecommand*\mciteBstWouldAddEndPuncttrue
  {\def\EndOfBibitem{\unskip.}}
\providecommand*\mciteBstWouldAddEndPunctfalse
  {\let\EndOfBibitem\relax}
\providecommand*\mciteSetBstMidEndSepPunct[3]{}
\providecommand*\mciteSetBstSublistLabelBeginEnd[3]{}
\providecommand*\EndOfBibitem{}
\mciteSetBstSublistMode{f}
\mciteSetBstMaxWidthForm{subitem}{(\alph{mcitesubitemcount})}
\mciteSetBstSublistLabelBeginEnd
  {\mcitemaxwidthsubitemform\space}
  {\relax}
  {\relax}

\bibitem[Szabo and Ostlund(1996)Szabo, and Ostlund]{SzaboBook}
Szabo,~A.; Ostlund,~N.~S. \emph{Modern Quantum Chemistry}; Dover Publications, 1996\relax
\mciteBstWouldAddEndPuncttrue
\mciteSetBstMidEndSepPunct{\mcitedefaultmidpunct}
{\mcitedefaultendpunct}{\mcitedefaultseppunct}\relax
\EndOfBibitem
\bibitem[Helgaker \latin{et~al.}(2000)Helgaker, J{\o}rgensen, and Olsen]{HelgakerBook}
Helgaker,~T.; J{\o}rgensen,~P.; Olsen,~J. \emph{Molecular Electronic Structure Theory}; John Wiley and Sons, 2000\relax
\mciteBstWouldAddEndPuncttrue
\mciteSetBstMidEndSepPunct{\mcitedefaultmidpunct}
{\mcitedefaultendpunct}{\mcitedefaultseppunct}\relax
\EndOfBibitem
\bibitem[Feynman(1982)]{Feynman1982}
Feynman,~R.~P. Simulating Physics with Computers. \emph{Int. J. Theor. Phys.} \textbf{1982}, \emph{21}, 467--488\relax
\mciteBstWouldAddEndPuncttrue
\mciteSetBstMidEndSepPunct{\mcitedefaultmidpunct}
{\mcitedefaultendpunct}{\mcitedefaultseppunct}\relax
\EndOfBibitem
\bibitem[McArdle \latin{et~al.}(2020)McArdle, Endo, Aspuru-Guzik, Benjamin, and Yuan]{McArdle2020}
McArdle,~S.; Endo,~S.; Aspuru-Guzik,~A.; Benjamin,~S.~C.; Yuan,~X. Quantum computational chemistry. \emph{Rev. Mod. Phys.} \textbf{2020}, \emph{92}, 015003\relax
\mciteBstWouldAddEndPuncttrue
\mciteSetBstMidEndSepPunct{\mcitedefaultmidpunct}
{\mcitedefaultendpunct}{\mcitedefaultseppunct}\relax
\EndOfBibitem
\bibitem[Peruzzo \latin{et~al.}(2014)Peruzzo, McClean, Shadbolt, Yung, Zhou, Love, Aspuru-Guzik, and O'Brien]{Peruzzo2014}
Peruzzo,~A.; McClean,~J.; Shadbolt,~P.; Yung,~M.-H.; Zhou,~X.-Q.; Love,~P.~J.; Aspuru-Guzik,~A.; O'Brien,~J.~L. A Variational Eigenvalue Solver on a Photonic Quantum Processor. \emph{Nat. Commun.} \textbf{2014}, \emph{5}, 4213\relax
\mciteBstWouldAddEndPuncttrue
\mciteSetBstMidEndSepPunct{\mcitedefaultmidpunct}
{\mcitedefaultendpunct}{\mcitedefaultseppunct}\relax
\EndOfBibitem
\bibitem[Nakanishi \latin{et~al.}(2019)Nakanishi, Mitarai, and Fujii]{Nakanishi_2019}
Nakanishi,~K.~M.; Mitarai,~K.; Fujii,~K. Subspace-search variational quantum eigensolver for excited states. \emph{Phys. Rev. Res.} \textbf{2019}, \emph{1}, 033062\relax
\mciteBstWouldAddEndPuncttrue
\mciteSetBstMidEndSepPunct{\mcitedefaultmidpunct}
{\mcitedefaultendpunct}{\mcitedefaultseppunct}\relax
\EndOfBibitem
\bibitem[McClean \latin{et~al.}(2017)McClean, Kimchi-Schwartz, Carter, and de~Jong]{McClean2017}
McClean,~J.~R.; Kimchi-Schwartz,~M.~E.; Carter,~J.; de~Jong,~W.~A. Hybrid quantum-classical hierarchy for mitigation of decoherence and determination of excited states. \emph{Phys. Rev. A} \textbf{2017}, \emph{95}, 042308\relax
\mciteBstWouldAddEndPuncttrue
\mciteSetBstMidEndSepPunct{\mcitedefaultmidpunct}
{\mcitedefaultendpunct}{\mcitedefaultseppunct}\relax
\EndOfBibitem
\bibitem[Higgott \latin{et~al.}(2019)Higgott, Wang, and Brierley]{Higgott_2019}
Higgott,~O.; Wang,~D.; Brierley,~S. Variational Quantum Computation of Excited States. \emph{Quantum} \textbf{2019}, \emph{3}, 156\relax
\mciteBstWouldAddEndPuncttrue
\mciteSetBstMidEndSepPunct{\mcitedefaultmidpunct}
{\mcitedefaultendpunct}{\mcitedefaultseppunct}\relax
\EndOfBibitem
\bibitem[Wang and Mazziotti(2023)Wang, and Mazziotti]{wang2023electronicexcitedstatesvariancebased}
Wang,~Y.; Mazziotti,~D.~A. {Electronic excited states from a variance-based contracted quantum eigensolver}. \emph{Phys. Rev. A} \textbf{2023}, \emph{108}, 022814\relax
\mciteBstWouldAddEndPuncttrue
\mciteSetBstMidEndSepPunct{\mcitedefaultmidpunct}
{\mcitedefaultendpunct}{\mcitedefaultseppunct}\relax
\EndOfBibitem
\bibitem[Reiher \latin{et~al.}(2017)Reiher, Wiebe, Svore, Wecker, and Troyer]{Reiher_2017}
Reiher,~M.; Wiebe,~N.; Svore,~K.~M.; Wecker,~D.; Troyer,~M. Elucidating reaction mechanisms on quantum computers. \emph{Proc. Natl. Acad. Sci. U.S.A.} \textbf{2017}, \emph{114}, 7555–7560\relax
\mciteBstWouldAddEndPuncttrue
\mciteSetBstMidEndSepPunct{\mcitedefaultmidpunct}
{\mcitedefaultendpunct}{\mcitedefaultseppunct}\relax
\EndOfBibitem
\bibitem[Mei and Yang(2019)Mei, and Yang]{Mei2019excited}
Mei,~Y.; Yang,~W. Excited-state potential energy surfaces, conical intersections, and analytical gradients from ground-state density functional theory. \emph{J. Phys. Chem. Lett.} \textbf{2019}, \emph{10}, 2538--2545\relax
\mciteBstWouldAddEndPuncttrue
\mciteSetBstMidEndSepPunct{\mcitedefaultmidpunct}
{\mcitedefaultendpunct}{\mcitedefaultseppunct}\relax
\EndOfBibitem
\bibitem[Lischka \latin{et~al.}(2018)Lischka, Nachtigallová, Aquino, Szalay, Plasser, Machado, and Barbatti]{Lischka2018}
Lischka,~H.; Nachtigallová,~D.; Aquino,~A. J.~A.; Szalay,~P.~G.; Plasser,~F.; Machado,~F. B.~C.; Barbatti,~M. Multireference Approaches for Excited States of Molecules. \emph{Chem. Rev.} \textbf{2018}, \emph{118}, 7293--7361\relax
\mciteBstWouldAddEndPuncttrue
\mciteSetBstMidEndSepPunct{\mcitedefaultmidpunct}
{\mcitedefaultendpunct}{\mcitedefaultseppunct}\relax
\EndOfBibitem
\bibitem[Cianci \latin{et~al.}(2024)Cianci, Santos, and Batista]{cianci2024subspacesearchquantumimaginarytime}
Cianci,~C.; Santos,~L.~F.; Batista,~V.~S. Subspace-Search Quantum Imaginary Time Evolution for Excited State Computations. \emph{J. Chem. Theory Comput.} \textbf{2024}, \emph{20}, 8940--8947\relax
\mciteBstWouldAddEndPuncttrue
\mciteSetBstMidEndSepPunct{\mcitedefaultmidpunct}
{\mcitedefaultendpunct}{\mcitedefaultseppunct}\relax
\EndOfBibitem
\bibitem[McClean \latin{et~al.}(2016)McClean, Romero, Babbush, and Aspuru-Guzik]{Mcclean2016theory}
McClean,~J.~R.; Romero,~J.; Babbush,~R.; Aspuru-Guzik,~A. The theory of variational hybrid quantum-classical algorithms. \emph{New J. Phys.} \textbf{2016}, \emph{18}, 023023\relax
\mciteBstWouldAddEndPuncttrue
\mciteSetBstMidEndSepPunct{\mcitedefaultmidpunct}
{\mcitedefaultendpunct}{\mcitedefaultseppunct}\relax
\EndOfBibitem
\bibitem[Baek \latin{et~al.}(2023)Baek, Hait, Shee, Leimkuhler, Huggins, Stetina, Head-Gordon, and Whaley]{Baek2023}
Baek,~U.; Hait,~D.; Shee,~J.; Leimkuhler,~O.; Huggins,~W.~J.; Stetina,~T.~F.; Head-Gordon,~M.; Whaley,~K.~B. Say {NO} to Optimization: {A} Nonorthogonal Quantum Eigensolver. \emph{PRX Quantum} \textbf{2023}, \emph{4}, 030307\relax
\mciteBstWouldAddEndPuncttrue
\mciteSetBstMidEndSepPunct{\mcitedefaultmidpunct}
{\mcitedefaultendpunct}{\mcitedefaultseppunct}\relax
\EndOfBibitem
\bibitem[Zheng \latin{et~al.}(2024)Zheng, Peng, Li, Yang, and Kowalski]{Zheng2024GCM}
Zheng,~M.; Peng,~B.; Li,~A.; Yang,~X.; Kowalski,~K. Unleashed from constrained optimization: {Q}uantum computing for quantum chemistry employing generator coordinate inspired method. \emph{npj Quantum Inf.} \textbf{2024}, \emph{10}, 127\relax
\mciteBstWouldAddEndPuncttrue
\mciteSetBstMidEndSepPunct{\mcitedefaultmidpunct}
{\mcitedefaultendpunct}{\mcitedefaultseppunct}\relax
\EndOfBibitem
\bibitem[Claudino \latin{et~al.}(2023)Claudino, Peng, Kowalski, and Humble]{Claudino2023modeling}
Claudino,~D.; Peng,~B.; Kowalski,~K.; Humble,~T.~S. Modeling singlet fission on a quantum computer. \emph{J. Phys. Chem. Lett.} \textbf{2023}, \emph{14}, 5511--5516\relax
\mciteBstWouldAddEndPuncttrue
\mciteSetBstMidEndSepPunct{\mcitedefaultmidpunct}
{\mcitedefaultendpunct}{\mcitedefaultseppunct}\relax
\EndOfBibitem
\bibitem[Dutta \latin{et~al.}(2025)Dutta, Vu, Xu, Cabral, Lyu, Soudackov, Dan, Li, Wang, and Batista]{Dutta2024EST}
Dutta,~R.; Vu,~N.~P.; Xu,~C.; Cabral,~D.~G.; Lyu,~N.; Soudackov,~A.~V.; Dan,~X.; Li,~H.; Wang,~C.; Batista,~V.~S. Simulating Electronic Structure on Bosonic Quantum Computers. \emph{J. Chem. Theory Comput.} \textbf{2025}, \emph{21}, 2281--2300\relax
\mciteBstWouldAddEndPuncttrue
\mciteSetBstMidEndSepPunct{\mcitedefaultmidpunct}
{\mcitedefaultendpunct}{\mcitedefaultseppunct}\relax
\EndOfBibitem
\bibitem[Dutta \latin{et~al.}(2025)Dutta, Allen, Vu, Xu, Liu, Miao, Wang, Surana, Wang, Ding, \latin{et~al.} others]{Dutta2025solving}
Dutta,~R.; Allen,~B.; Vu,~N.~P.; Xu,~C.; Liu,~K.; Miao,~F.; Wang,~B.; Surana,~A.; Wang,~C.; Ding,~Y.; others Solving Constrained Optimization Problems Using Hybrid Qubit-Qumode Quantum Devices. \emph{arXiv preprint arXiv:2501.11735} \textbf{2025}, \relax
\mciteBstWouldAddEndPunctfalse
\mciteSetBstMidEndSepPunct{\mcitedefaultmidpunct}
{}{\mcitedefaultseppunct}\relax
\EndOfBibitem
\bibitem[Zhang and Zhuang(2024)Zhang, and Zhuang]{Zhang2024energy}
Zhang,~B.; Zhuang,~Q. Energy-dependent barren plateau in bosonic variational quantum circuits. \emph{Quantum Sci. Technol.} \textbf{2024}, \emph{10}, 015009\relax
\mciteBstWouldAddEndPuncttrue
\mciteSetBstMidEndSepPunct{\mcitedefaultmidpunct}
{\mcitedefaultendpunct}{\mcitedefaultseppunct}\relax
\EndOfBibitem
\bibitem[Kan \latin{et~al.}(2024)Kan, Palma, Du, Stein, Liu, Chen, Li, and Mao]{Kan2024}
Kan,~S.; Palma,~M.; Du,~Z.; Stein,~S.~A.; Liu,~C.; Chen,~J.; Li,~A.; Mao,~Y. Benchmarking Optimizers for Qumode State Preparation with Variational Quantum Algorithms. 2024 IEEE International Conference on Quantum Computing and Engineering (QCE). 2024; pp 1558--1564\relax
\mciteBstWouldAddEndPuncttrue
\mciteSetBstMidEndSepPunct{\mcitedefaultmidpunct}
{\mcitedefaultendpunct}{\mcitedefaultseppunct}\relax
\EndOfBibitem
\bibitem[Liao \latin{et~al.}(2024)Liao, Zhang, and Zhuang]{Liao2024quantum}
Liao,~P.; Zhang,~B.; Zhuang,~Q. Quantum-enhanced learning with a controllable bosonic variational sensor network. \emph{Quantum Sci. Technol.} \textbf{2024}, \emph{9}, 045040\relax
\mciteBstWouldAddEndPuncttrue
\mciteSetBstMidEndSepPunct{\mcitedefaultmidpunct}
{\mcitedefaultendpunct}{\mcitedefaultseppunct}\relax
\EndOfBibitem
\bibitem[Araz \latin{et~al.}(2025)Araz, Grau, Montgomery, and Ringer]{Araz2025hybrid}
Araz,~J.~Y.; Grau,~M.; Montgomery,~J.; Ringer,~F. Hybrid quantum simulations with qubits and qumodes on trapped-ion platforms. \emph{Phys. Rev. A} \textbf{2025}, \emph{112}, 012620\relax
\mciteBstWouldAddEndPuncttrue
\mciteSetBstMidEndSepPunct{\mcitedefaultmidpunct}
{\mcitedefaultendpunct}{\mcitedefaultseppunct}\relax
\EndOfBibitem
\bibitem[Dutta \latin{et~al.}(2024)Dutta, Cabral, Lyu, Vu, Wang, Allen, Dan, Corti{\~n}as, Khazaei, Smart, \latin{et~al.} others]{Dutta2024perspective}
Dutta,~R.; Cabral,~D.~G.; Lyu,~N.; Vu,~N.~P.; Wang,~Y.; Allen,~B.; Dan,~X.; Corti{\~n}as,~R.~G.; Khazaei,~P.; Smart,~S.~E.; others Simulating Chemistry on Bosonic Quantum Devices. \emph{J. Chem. Theory Comput.} \textbf{2024}, \emph{20}, 6426\relax
\mciteBstWouldAddEndPuncttrue
\mciteSetBstMidEndSepPunct{\mcitedefaultmidpunct}
{\mcitedefaultendpunct}{\mcitedefaultseppunct}\relax
\EndOfBibitem
\bibitem[Blais \latin{et~al.}(2021)Blais, Grimsmo, Girvin, and Wallraff]{Blais2021}
Blais,~A.; Grimsmo,~A.~L.; Girvin,~S.~M.; Wallraff,~A. Circuit quantum electrodynamics. \emph{Rev. Mod. Phys.} \textbf{2021}, \emph{93}, 025005\relax
\mciteBstWouldAddEndPuncttrue
\mciteSetBstMidEndSepPunct{\mcitedefaultmidpunct}
{\mcitedefaultendpunct}{\mcitedefaultseppunct}\relax
\EndOfBibitem
\bibitem[Liu \latin{et~al.}(2024)Liu, Singh, Smith, Crane, Martyn, Eickbusch, Schuckert, Li, Sinanan-Singh, Soley, \latin{et~al.} others]{Liu2024qumodequbitreview}
Liu,~Y.; Singh,~S.; Smith,~K.~C.; Crane,~E.; Martyn,~J.~M.; Eickbusch,~A.; Schuckert,~A.; Li,~R.~D.; Sinanan-Singh,~J.; Soley,~M.~B.; others Hybrid oscillator-qubit quantum processors: Instruction set architectures, abstract machine models, and applications. \emph{arXiv preprint arXiv:2407.10381} \textbf{2024}, \relax
\mciteBstWouldAddEndPunctfalse
\mciteSetBstMidEndSepPunct{\mcitedefaultmidpunct}
{}{\mcitedefaultseppunct}\relax
\EndOfBibitem
\bibitem[Copetudo \latin{et~al.}(2024)Copetudo, Fontaine, Valadares, and Gao]{Copetudo2024}
Copetudo,~A.; Fontaine,~C.~Y.; Valadares,~F.; Gao,~Y.~Y. Shaping photons: {Quantum} information processing with bosonic {cQED}. \emph{Appl. Phys. Lett.} \textbf{2024}, \emph{124}, 080502\relax
\mciteBstWouldAddEndPuncttrue
\mciteSetBstMidEndSepPunct{\mcitedefaultmidpunct}
{\mcitedefaultendpunct}{\mcitedefaultseppunct}\relax
\EndOfBibitem
\bibitem[Kudra \latin{et~al.}(2022)Kudra, Kervinen, Strandberg, Ahmed, Scigliuzzo, Osman, Lozano, Thol{\'e}n, Borgani, Haviland, \latin{et~al.} others]{Kudra2022}
Kudra,~M.; Kervinen,~M.; Strandberg,~I.; Ahmed,~S.; Scigliuzzo,~M.; Osman,~A.; Lozano,~D.~P.; Thol{\'e}n,~M.~O.; Borgani,~R.; Haviland,~D.~B.; others Robust preparation of Wigner-negative states with optimized {SNAP}-displacement sequences. \emph{PRX Quantum} \textbf{2022}, \emph{3}, 030301\relax
\mciteBstWouldAddEndPuncttrue
\mciteSetBstMidEndSepPunct{\mcitedefaultmidpunct}
{\mcitedefaultendpunct}{\mcitedefaultseppunct}\relax
\EndOfBibitem
\bibitem[Job(2023)]{Job2023efficient}
Job,~J. Efficient, direct compilation of SU (N) operations into SNAP \& Displacement gates. \emph{arXiv preprint arXiv:2307.11900} \textbf{2023}, \relax
\mciteBstWouldAddEndPunctfalse
\mciteSetBstMidEndSepPunct{\mcitedefaultmidpunct}
{}{\mcitedefaultseppunct}\relax
\EndOfBibitem
\bibitem[Krastanov \latin{et~al.}(2015)Krastanov, Albert, Shen, Zou, Heeres, Vlastakis, Schoelkopf, and Jiang]{Krastanov2015}
Krastanov,~S.; Albert,~V.~V.; Shen,~C.; Zou,~C.-L.; Heeres,~R.~W.; Vlastakis,~B.; Schoelkopf,~R.~J.; Jiang,~L. Universal control of an oscillator with dispersive coupling to a qubit. \emph{Phys. Rev. A} \textbf{2015}, \emph{92}, 040303\relax
\mciteBstWouldAddEndPuncttrue
\mciteSetBstMidEndSepPunct{\mcitedefaultmidpunct}
{\mcitedefaultendpunct}{\mcitedefaultseppunct}\relax
\EndOfBibitem
\bibitem[Huang \latin{et~al.}(2025)Huang, DiNapoli, Rockwood, Yuan, Narasimhan, Gupta, Bal, Crisa, Garattoni, Lu, \latin{et~al.} others]{Huang2025fast}
Huang,~J.; DiNapoli,~T.~J.; Rockwood,~G.; Yuan,~M.; Narasimhan,~P.; Gupta,~E.; Bal,~M.; Crisa,~F.; Garattoni,~S.; Lu,~Y.; others Fast Sideband Control of a Weakly Coupled Multimode Bosonic Memory. \emph{arXiv preprint arXiv:2503.10623} \textbf{2025}, \relax
\mciteBstWouldAddEndPunctfalse
\mciteSetBstMidEndSepPunct{\mcitedefaultmidpunct}
{}{\mcitedefaultseppunct}\relax
\EndOfBibitem
\bibitem[Jordan(1935)]{Jordan1935}
Jordan,~P. Der Zusammenhang der symmetrischen und linearen Gruppen und das Mehrk{\"o}rperproblem. \emph{Zeitschrift f{\"u}r Physik} \textbf{1935}, \emph{94}, 531\relax
\mciteBstWouldAddEndPuncttrue
\mciteSetBstMidEndSepPunct{\mcitedefaultmidpunct}
{\mcitedefaultendpunct}{\mcitedefaultseppunct}\relax
\EndOfBibitem
\bibitem[McArdle \latin{et~al.}(2019)McArdle, Jones, Endo, Li, Benjamin, and Yuan]{McArdle_2019}
McArdle,~S.; Jones,~T.; Endo,~S.; Li,~Y.; Benjamin,~S.~C.; Yuan,~X. Variational ansatz-based quantum simulation of imaginary time evolution. \emph{npj Quantum Inf.} \textbf{2019}, \emph{5}, 75\relax
\mciteBstWouldAddEndPuncttrue
\mciteSetBstMidEndSepPunct{\mcitedefaultmidpunct}
{\mcitedefaultendpunct}{\mcitedefaultseppunct}\relax
\EndOfBibitem
\bibitem[Tsuchimochi \latin{et~al.}(2023)Tsuchimochi, Ryo, Ten-no, and Sasasako]{doi:10.1021/acs.jctc.2c00906}
Tsuchimochi,~T.; Ryo,~Y.; Ten-no,~S.~L.; Sasasako,~K. Improved Algorithms of Quantum Imaginary Time Evolution for Ground and Excited States of Molecular Systems. \emph{J. Chem. Theory Comput.} \textbf{2023}, \emph{19}, 503--513\relax
\mciteBstWouldAddEndPuncttrue
\mciteSetBstMidEndSepPunct{\mcitedefaultmidpunct}
{\mcitedefaultendpunct}{\mcitedefaultseppunct}\relax
\EndOfBibitem
\bibitem[Smart and Mazziotti(2021)Smart, and Mazziotti]{Smart2021}
Smart,~S.~E.; Mazziotti,~D.~A. Quantum solver of contracted eigenvalue equations for scalable molecular simulations on quantum computing devices. \emph{Phys. Rev. Lett.} \textbf{2021}, \emph{126}, 070504\relax
\mciteBstWouldAddEndPuncttrue
\mciteSetBstMidEndSepPunct{\mcitedefaultmidpunct}
{\mcitedefaultendpunct}{\mcitedefaultseppunct}\relax
\EndOfBibitem
\bibitem[Smart \latin{et~al.}(2024)Smart, Welakuh, and Narang]{Smart_2024}
Smart,~S.~E.; Welakuh,~D.~M.; Narang,~P. Many-Body Excited States with a Contracted Quantum Eigensolver. \emph{J. Chem. Theory Comput.} \textbf{2024}, \emph{20}, 3580–3589\relax
\mciteBstWouldAddEndPuncttrue
\mciteSetBstMidEndSepPunct{\mcitedefaultmidpunct}
{\mcitedefaultendpunct}{\mcitedefaultseppunct}\relax
\EndOfBibitem
\bibitem[Benavides-Riveros \latin{et~al.}(2024)Benavides-Riveros, Wang, Warren, and Mazziotti]{Benavides_Riveros_2024}
Benavides-Riveros,~C.~L.; Wang,~Y.; Warren,~S.; Mazziotti,~D.~A. Quantum simulation of excited states from parallel contracted quantum eigensolvers. \emph{New J. Phys.} \textbf{2024}, \emph{26}, 033020\relax
\mciteBstWouldAddEndPuncttrue
\mciteSetBstMidEndSepPunct{\mcitedefaultmidpunct}
{\mcitedefaultendpunct}{\mcitedefaultseppunct}\relax
\EndOfBibitem
\bibitem[Colless \latin{et~al.}(2018)Colless, Ramasesh, Dahlen, Blok, Kimchi-Schwartz, McClean, Carter, de~Jong, and Siddiqi]{Colless2018spectra}
Colless,~J.~I.; Ramasesh,~V.~V.; Dahlen,~D.; Blok,~M.~S.; Kimchi-Schwartz,~M.~E.; McClean,~J.~R.; Carter,~J.; de~Jong,~W.~A.; Siddiqi,~I. Computation of Molecular Spectra on a Quantum Processor with an Error-Resilient Algorithm. \emph{Phys. Rev. X} \textbf{2018}, \emph{8}, 011021\relax
\mciteBstWouldAddEndPuncttrue
\mciteSetBstMidEndSepPunct{\mcitedefaultmidpunct}
{\mcitedefaultendpunct}{\mcitedefaultseppunct}\relax
\EndOfBibitem
\bibitem[Motta \latin{et~al.}(2019)Motta, Sun, Tan, O’Rourke, Ye, Minnich, Brandão, and Chan]{Motta_2019}
Motta,~M.; Sun,~C.; Tan,~A. T.~K.; O’Rourke,~M.~J.; Ye,~E.; Minnich,~A.~J.; Brandão,~F. G. S.~L.; Chan,~G. K.-L. Determining eigenstates and thermal states on a quantum computer using quantum imaginary time evolution. \emph{Nat. Phys.} \textbf{2019}, \emph{16}, 205–210\relax
\mciteBstWouldAddEndPuncttrue
\mciteSetBstMidEndSepPunct{\mcitedefaultmidpunct}
{\mcitedefaultendpunct}{\mcitedefaultseppunct}\relax
\EndOfBibitem
\bibitem[Tkachenko \latin{et~al.}(2023)Tkachenko, Cincio, Boldyrev, Tretiak, Dub, and Zhang]{tkachenko2023quantumdavidsonalgorithmexcited}
Tkachenko,~N.~V.; Cincio,~L.; Boldyrev,~A.~I.; Tretiak,~S.; Dub,~P.~A.; Zhang,~Y. Quantum Davidson Algorithm for Excited States. 2023; \url{https://arxiv.org/abs/2204.10741}\relax
\mciteBstWouldAddEndPuncttrue
\mciteSetBstMidEndSepPunct{\mcitedefaultmidpunct}
{\mcitedefaultendpunct}{\mcitedefaultseppunct}\relax
\EndOfBibitem
\bibitem[Lenzini \latin{et~al.}(2018)Lenzini, Janousek, Thearle, Villa, Haylock, Kasture, Cui, Phan, Dao, Yonezawa, \latin{et~al.} others]{Lenzini2018integrated}
Lenzini,~F.; Janousek,~J.; Thearle,~O.; Villa,~M.; Haylock,~B.; Kasture,~S.; Cui,~L.; Phan,~H.-P.; Dao,~D.~V.; Yonezawa,~H.; others Integrated photonic platform for quantum information with continuous variables. \emph{Sci. Adv.} \textbf{2018}, \emph{4}, eaat9331\relax
\mciteBstWouldAddEndPuncttrue
\mciteSetBstMidEndSepPunct{\mcitedefaultmidpunct}
{\mcitedefaultendpunct}{\mcitedefaultseppunct}\relax
\EndOfBibitem
\bibitem[Sarma \latin{et~al.}(2021)Sarma, Chakraborty, and Kalita]{Sarma2021continuous}
Sarma,~A.~K.; Chakraborty,~S.; Kalita,~S. Continuous variable quantum entanglement in optomechanical systems: A short review. \emph{AVS Quantum Sci.} \textbf{2021}, \emph{3}, 015901\relax
\mciteBstWouldAddEndPuncttrue
\mciteSetBstMidEndSepPunct{\mcitedefaultmidpunct}
{\mcitedefaultendpunct}{\mcitedefaultseppunct}\relax
\EndOfBibitem
\bibitem[Wang \latin{et~al.}(2020)Wang, Curtis, Lester, Zhang, Gao, Freeze, Batista, Vaccaro, Chuang, Frunzio, Jiang, Girvin, and Schoelkopf]{Wang2020vibronic}
Wang,~C.~S.; Curtis,~J.~C.; Lester,~B.~J.; Zhang,~Y.; Gao,~Y.~Y.; Freeze,~J.; Batista,~V.~S.; Vaccaro,~P.~H.; Chuang,~I.~L.; Frunzio,~L.; Jiang,~L.; Girvin,~S.~M.; Schoelkopf,~R.~J. Efficient Multiphoton Sampling of Molecular Vibronic Spectra on a Superconducting Bosonic Processor. \emph{Phys. Rev. X} \textbf{2020}, \emph{10}, 021060\relax
\mciteBstWouldAddEndPuncttrue
\mciteSetBstMidEndSepPunct{\mcitedefaultmidpunct}
{\mcitedefaultendpunct}{\mcitedefaultseppunct}\relax
\EndOfBibitem
\bibitem[Eickbusch \latin{et~al.}(2022)Eickbusch, Sivak, Ding, Elder, Jha, Venkatraman, Royer, Girvin, Schoelkopf, and Devoret]{Eickbusch2022}
Eickbusch,~A.; Sivak,~V.; Ding,~A.~Z.; Elder,~S.~S.; Jha,~S.~R.; Venkatraman,~J.; Royer,~B.; Girvin,~S.~M.; Schoelkopf,~R.~J.; Devoret,~M.~H. Fast universal control of an oscillator with weak dispersive coupling to a qubit. \emph{Nat. Phys.} \textbf{2022}, \emph{18}, 1464\relax
\mciteBstWouldAddEndPuncttrue
\mciteSetBstMidEndSepPunct{\mcitedefaultmidpunct}
{\mcitedefaultendpunct}{\mcitedefaultseppunct}\relax
\EndOfBibitem
\bibitem[Hillmann \latin{et~al.}(2020)Hillmann, Quijandr\'{\i}a, Johansson, Ferraro, Gasparinetti, and Ferrini]{Hillmann2020}
Hillmann,~T.; Quijandr\'{\i}a,~F.; Johansson,~G.; Ferraro,~A.; Gasparinetti,~S.; Ferrini,~G. Universal Gate Set for Continuous-Variable Quantum Computation with Microwave Circuits. \emph{Phys. Rev. Lett.} \textbf{2020}, \emph{125}, 160501\relax
\mciteBstWouldAddEndPuncttrue
\mciteSetBstMidEndSepPunct{\mcitedefaultmidpunct}
{\mcitedefaultendpunct}{\mcitedefaultseppunct}\relax
\EndOfBibitem
\bibitem[F{\"o}sel \latin{et~al.}(2020)F{\"o}sel, Krastanov, Marquardt, and Jiang]{Fosel2020}
F{\"o}sel,~T.; Krastanov,~S.; Marquardt,~F.; Jiang,~L. Efficient cavity control with {SNAP} gates. \emph{arXiv preprint arXiv:2004.14256} \textbf{2020}, \relax
\mciteBstWouldAddEndPunctfalse
\mciteSetBstMidEndSepPunct{\mcitedefaultmidpunct}
{}{\mcitedefaultseppunct}\relax
\EndOfBibitem
\bibitem[Curtis \latin{et~al.}(2021)Curtis, Hann, Elder, Wang, Frunzio, Jiang, and Schoelkopf]{Curtis2021}
Curtis,~J.~C.; Hann,~C.~T.; Elder,~S.~S.; Wang,~C.~S.; Frunzio,~L.; Jiang,~L.; Schoelkopf,~R.~J. Single-shot number-resolved detection of microwave photons with error mitigation. \emph{Phys. Rev. A} \textbf{2021}, \emph{103}, 023705\relax
\mciteBstWouldAddEndPuncttrue
\mciteSetBstMidEndSepPunct{\mcitedefaultmidpunct}
{\mcitedefaultendpunct}{\mcitedefaultseppunct}\relax
\EndOfBibitem
\bibitem[Deng \latin{et~al.}(2024)Deng, Li, Chen, Ni, Cai, Mai, Zhang, Zheng, Yu, Zou, \latin{et~al.} others]{Deng2024quantum}
Deng,~X.; Li,~S.; Chen,~Z.-J.; Ni,~Z.; Cai,~Y.; Mai,~J.; Zhang,~L.; Zheng,~P.; Yu,~H.; Zou,~C.-L.; others Quantum-enhanced metrology with large Fock states. \emph{Nat. Phys.} \textbf{2024}, \emph{20}, 1874--1880\relax
\mciteBstWouldAddEndPuncttrue
\mciteSetBstMidEndSepPunct{\mcitedefaultmidpunct}
{\mcitedefaultendpunct}{\mcitedefaultseppunct}\relax
\EndOfBibitem
\bibitem[Mitarai \latin{et~al.}(2018)Mitarai, Negoro, Kitagawa, and Fujii]{Mitarai2018}
Mitarai,~K.; Negoro,~M.; Kitagawa,~M.; Fujii,~K. Quantum circuit learning. \emph{Phys. Rev. A} \textbf{2018}, \emph{98}, 032309\relax
\mciteBstWouldAddEndPuncttrue
\mciteSetBstMidEndSepPunct{\mcitedefaultmidpunct}
{\mcitedefaultendpunct}{\mcitedefaultseppunct}\relax
\EndOfBibitem
\bibitem[Schuld \latin{et~al.}(2019)Schuld, Bergholm, Gogolin, Izaac, and Killoran]{Schuld2019}
Schuld,~M.; Bergholm,~V.; Gogolin,~C.; Izaac,~J.; Killoran,~N. Evaluating analytic gradients on quantum hardware. \emph{Phys. Rev. A} \textbf{2019}, \emph{99}, 032331\relax
\mciteBstWouldAddEndPuncttrue
\mciteSetBstMidEndSepPunct{\mcitedefaultmidpunct}
{\mcitedefaultendpunct}{\mcitedefaultseppunct}\relax
\EndOfBibitem
\bibitem[Ma \latin{et~al.}(2021)Ma, Puri, Schoelkopf, Devoret, Girvin, and Jiang]{Ma2021quantum}
Ma,~W.-L.; Puri,~S.; Schoelkopf,~R.~J.; Devoret,~M.~H.; Girvin,~S.~M.; Jiang,~L. Quantum control of bosonic modes with superconducting circuits. \emph{Sci. Bull.} \textbf{2021}, \emph{66}, 1789--1805\relax
\mciteBstWouldAddEndPuncttrue
\mciteSetBstMidEndSepPunct{\mcitedefaultmidpunct}
{\mcitedefaultendpunct}{\mcitedefaultseppunct}\relax
\EndOfBibitem
\bibitem[Zhang and Jing(2024)Zhang, and Jing]{Zhang2024generating}
Zhang,~C.-y.; Jing,~J. Generating Fock-state superpositions from coherent states by selective measurement. \emph{Phys. Rev. A} \textbf{2024}, \emph{110}, 042421\relax
\mciteBstWouldAddEndPuncttrue
\mciteSetBstMidEndSepPunct{\mcitedefaultmidpunct}
{\mcitedefaultendpunct}{\mcitedefaultseppunct}\relax
\EndOfBibitem
\bibitem[Law and Eberly(1996)Law, and Eberly]{LawEberly1996}
Law,~C.~K.; Eberly,~J.~H. Arbitrary Control of a Quantum Electromagnetic Field. \emph{Phys. Rev. Lett.} \textbf{1996}, \emph{76}, 1055--1058\relax
\mciteBstWouldAddEndPuncttrue
\mciteSetBstMidEndSepPunct{\mcitedefaultmidpunct}
{\mcitedefaultendpunct}{\mcitedefaultseppunct}\relax
\EndOfBibitem
\bibitem[Mischuck and M\o{}lmer(2013)Mischuck, and M\o{}lmer]{Mischuck2013}
Mischuck,~B.; M\o{}lmer,~K. Qudit quantum computation in the Jaynes-Cummings model. \emph{Phys. Rev. A} \textbf{2013}, \emph{87}, 022341\relax
\mciteBstWouldAddEndPuncttrue
\mciteSetBstMidEndSepPunct{\mcitedefaultmidpunct}
{\mcitedefaultendpunct}{\mcitedefaultseppunct}\relax
\EndOfBibitem
\bibitem[Liu \latin{et~al.}(2021)Liu, Sinanan-Singh, Kearney, Mintzer, and Chuang]{Liu2021qudit}
Liu,~Y.; Sinanan-Singh,~J.; Kearney,~M.~T.; Mintzer,~G.; Chuang,~I.~L. Constructing qudits from infinite-dimensional oscillators by coupling to qubits. \emph{Phys. Rev. A} \textbf{2021}, \emph{104}, 032605\relax
\mciteBstWouldAddEndPuncttrue
\mciteSetBstMidEndSepPunct{\mcitedefaultmidpunct}
{\mcitedefaultendpunct}{\mcitedefaultseppunct}\relax
\EndOfBibitem
\bibitem[Virtanen \latin{et~al.}(2020)Virtanen, Gommers, Oliphant, Haberland, Reddy, Cournapeau, Burovski, Peterson, Weckesser, Bright, {van der Walt}, Brett, Wilson, Millman, Mayorov, Nelson, Jones, Kern, Larson, Carey, Polat, Feng, Moore, {VanderPlas}, Laxalde, Perktold, Cimrman, Henriksen, Quintero, Harris, Archibald, Ribeiro, Pedregosa, {van Mulbregt}, and {SciPy 1.0 Contributors}]{2020SciPy-NMeth}
Virtanen,~P.; Gommers,~R.; Oliphant,~T.~E.; Haberland,~M.; Reddy,~T.; Cournapeau,~D.; Burovski,~E.; Peterson,~P.; Weckesser,~W.; Bright,~J.; {van der Walt},~S.~J.; Brett,~M.; Wilson,~J.; Millman,~K.~J.; Mayorov,~N.; Nelson,~A. R.~J.; Jones,~E.; Kern,~R.; Larson,~E.; Carey,~C.~J.; Polat,~{\.I}.; Feng,~Y.; Moore,~E.~W.; {VanderPlas},~J.; Laxalde,~D.; Perktold,~J.; Cimrman,~R.; Henriksen,~I.; Quintero,~E.~A.; Harris,~C.~R.; Archibald,~A.~M.; Ribeiro,~A.~H.; Pedregosa,~F.; {van Mulbregt},~P.; {SciPy 1.0 Contributors} {{SciPy} 1.0: Fundamental Algorithms for Scientific Computing in Python}. \emph{Nat. Methods} \textbf{2020}, \emph{17}, 261--272\relax
\mciteBstWouldAddEndPuncttrue
\mciteSetBstMidEndSepPunct{\mcitedefaultmidpunct}
{\mcitedefaultendpunct}{\mcitedefaultseppunct}\relax
\EndOfBibitem
\bibitem[Wang \latin{et~al.}(2025)Wang, Cianci, Avdic, Dutta, Warren, Allen, Vu, Santos, Batista, and Mazziotti]{Wang2025characterizing}
Wang,~Y.; Cianci,~C.; Avdic,~I.; Dutta,~R.; Warren,~S.; Allen,~B.; Vu,~N.~P.; Santos,~L.~F.; Batista,~V.~S.; Mazziotti,~D.~A. Characterizing conical intersections of nucleobases on quantum computers. \emph{J. Chem. Theory Comput.} \textbf{2025}, \emph{21}, 1213\relax
\mciteBstWouldAddEndPuncttrue
\mciteSetBstMidEndSepPunct{\mcitedefaultmidpunct}
{\mcitedefaultendpunct}{\mcitedefaultseppunct}\relax
\EndOfBibitem
\bibitem[Wang \latin{et~al.}(2023)Wang, Sager-Smith, and Mazziotti]{Wang2023quantum}
Wang,~Y.; Sager-Smith,~L.~M.; Mazziotti,~D.~A. Quantum simulation of bosons with the contracted quantum eigensolver. \emph{New J. Phys.} \textbf{2023}, \emph{25}, 103005\relax
\mciteBstWouldAddEndPuncttrue
\mciteSetBstMidEndSepPunct{\mcitedefaultmidpunct}
{\mcitedefaultendpunct}{\mcitedefaultseppunct}\relax
\EndOfBibitem
\bibitem[Peng \latin{et~al.}(2025)Peng, Su, Claudino, Kowalski, Low, and Roetteler]{Peng2025quantum}
Peng,~B.; Su,~Y.; Claudino,~D.; Kowalski,~K.; Low,~G.~H.; Roetteler,~M. Quantum simulation of boson-related hamiltonians: techniques, effective hamiltonian construction, and error analysis. \emph{Quantum Sci. Technol.} \textbf{2025}, \emph{10}, 023002\relax
\mciteBstWouldAddEndPuncttrue
\mciteSetBstMidEndSepPunct{\mcitedefaultmidpunct}
{\mcitedefaultendpunct}{\mcitedefaultseppunct}\relax
\EndOfBibitem
\bibitem[Barone \latin{et~al.}(2012)Barone, Biczysko, Bloino, Borkowska-Panek, Carnimeo, and Panek]{Barone2012anharmonic}
Barone,~V.; Biczysko,~M.; Bloino,~J.; Borkowska-Panek,~M.; Carnimeo,~I.; Panek,~P. Toward anharmonic computations of vibrational spectra for large molecular systems. \emph{Int. J. Quantum Chem.} \textbf{2012}, \emph{112}, 2185--2200\relax
\mciteBstWouldAddEndPuncttrue
\mciteSetBstMidEndSepPunct{\mcitedefaultmidpunct}
{\mcitedefaultendpunct}{\mcitedefaultseppunct}\relax
\EndOfBibitem
\bibitem[Wang \latin{et~al.}(2022)Wang, Ren, Li, and Shuai]{Wang2022anharmonic}
Wang,~Y.; Ren,~J.; Li,~W.; Shuai,~Z. Hybrid quantum-classical boson sampling algorithm for molecular vibrationally resolved electronic spectroscopy with Duschinsky rotation and anharmonicity. \emph{J. Phys. Chem. Lett} \textbf{2022}, \emph{13}, 6391--6399\relax
\mciteBstWouldAddEndPuncttrue
\mciteSetBstMidEndSepPunct{\mcitedefaultmidpunct}
{\mcitedefaultendpunct}{\mcitedefaultseppunct}\relax
\EndOfBibitem
\bibitem[You \latin{et~al.}(2024)You, Lu, Kim, Kurkcuoglu, Zhu, van Zanten, Roy, Lu, Chakram, Grassellino, Romanenko, Koch, and Zorzetti]{You2024Crosstalk}
You,~X.; Lu,~Y.; Kim,~T.; Kurkcuoglu,~D.~M.; Zhu,~S.; van Zanten,~D.; Roy,~T.; Lu,~Y.; Chakram,~S.; Grassellino,~A.; Romanenko,~A.; Koch,~J.; Zorzetti,~S. Crosstalk-Robust Quantum Control in Multimode Bosonic Systems. \emph{arXiv preprint arXiv:2403.00275} \textbf{2024}, \relax
\mciteBstWouldAddEndPunctfalse
\mciteSetBstMidEndSepPunct{\mcitedefaultmidpunct}
{}{\mcitedefaultseppunct}\relax
\EndOfBibitem
\end{mcitethebibliography}


\begin{tocentry}
\begin{center}

\includegraphics[width=0.9\columnwidth]{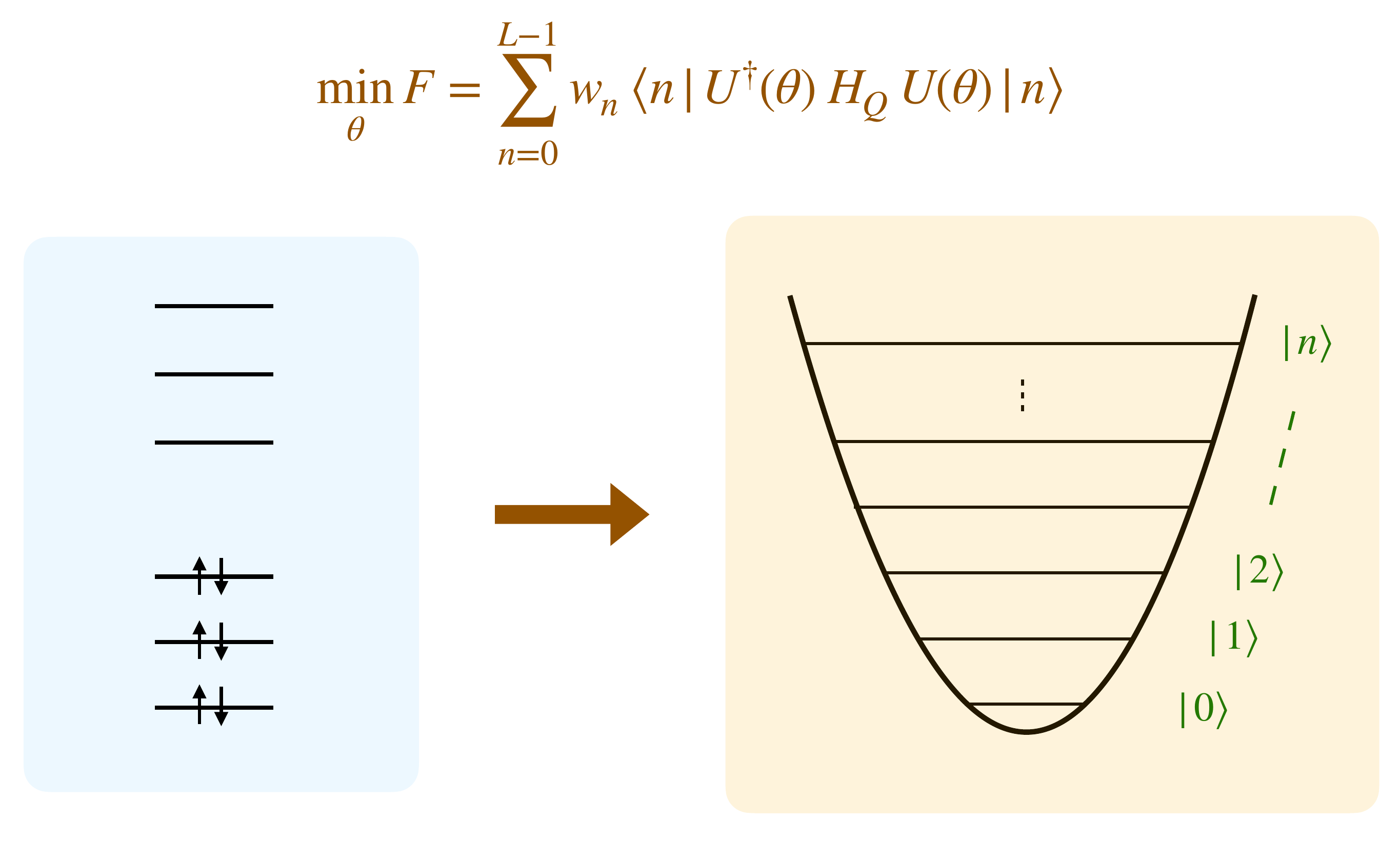}
    
\end{center}
\end{tocentry}


\end{document}